%% file: main.tex
\documentclass[hidelinks,onefignum,onetabnum]{siamart251216}

\input{shared}

\ifpdf
\hypersetup{
  pdftitle={Coarse-Graining},
  pdfauthor={W. J. M. Ridgway, and R. J. Spiteri}
}
\fi

\title{Coarse-Graining Agent-Based Models of Bacterial Infections}

\author{Wesley J. M. Ridgway\thanks{Department of Mathematics and Statistics,
University of Saskatchewan, Saskatoon, Saskatchewan, Canada. This author's research was supported in part by the Pacific Institute for the Mathematical Sciences.} 
\and
Raymond J. Spiteri\thanks{
Department of Computer Science, University of Saskatchewan\, Saskatoon, Saskatchewan, Canada. This author's research was supported in part by the Natural Sciences and Engineering Research Council of Canada.}}

\def\mbf#1{{\mathbf{#1}}}
\def\pfo#1#2{\frac{\partial#1}{\partial#2}}
\def\pft#1#2{\frac{\partial^2#1}{\partial#2^2}}
\def\dfo#1#2{\frac{\mathrm d {#1}}{\mathrm d {#2}}}
\def\dft#1#2{\frac{\mathrm d ^2{#1}}{\mathrm d {#2} ^2}}
\def\est{\mathrm{e.s.t.}}
\newcommand{\mtb}{Mtb}

\begin{document}

\maketitle

\begin{abstract}
Agent-based models (ABMs) provide a natural framework for representing
cell-level rules and spatial heterogeneity in bacterial infections, but their
computational cost limits their use for macroscopic tissue-scale simulations
and broad parameter exploration. We derive a deterministic coarse-grained
description for a class of bacterial-infection ABMs in which immune cells and
extracellular bacteria diffuse, immune cells ingest nearby bacteria, and
intracellular bacterial loads evolve through prescribed birth and clearance
processes. The first coarse-grained model is a semidiscrete
reaction--diffusion system that retains a discrete internal state for each
immune-cell bacterial load while representing cell and bacterial populations by
continuum concentration fields. The key technical step is the derivation of
state-dependent effective ingestion rates from the microscopic ABM parameters:
these rates are obtained by solving an auxiliary diffusion problem around a
single bacterium and computing the flux of immune cells into the interaction
region. We then take a continuum limit in the internal state variable, yielding
a state-structured reaction--diffusion system in which intracellular dynamics
appear as advection and diffusion in state space. Numerical comparisons with
ensemble-averaged ABM simulations show close agreement in biologically motivated
parameter regimes. The resulting framework preserves the rule-based structure
of the ABM while producing PDE models that are substantially more tractable for
large-scale simulation and parameter studies.
\end{abstract}

\section{Introduction}

Mathematical modelling has been a powerful tool for understanding the pathology
of bacterial infections. Within the past two decades, agent-based models (ABMs)
have become a popular approach for modelling within-host infections for two main
reasons. First, ABMs are attractive because the behaviour of cells and bacteria
can be translated naturally into rules. Second, ABMs can represent cellular
heterogeneity directly. Examples include ABMs of \emph{Mycobacterium
  tuberculosis} (\mtb) infections \cite{segovia2004identifying, petrucciani2024silico, ray2009synergy},
\emph{Leishmania major} infections \cite{dancik2010parameter}, bone and joint
infections \cite{alsassa2020modeling}, surgical site infections
\cite{gopalakrishnan2013using}, and \emph{Francisella tularensis} infections \cite{carruthers2020stochastic, gillard2014modeling}.  The main drawback of ABMs is
their computational cost, which often scales with the number of agents \(M\), or
possibly with \(M^2\) when pairwise interactions must be resolved. Even with
modern high-performance computing, simulating a macroscopic system, such as an organ, is
computationally infeasible. Moreover, because ABMs are often stochastic, many
independent realizations are needed to generate meaningful statistics.

In contrast, PDE models are relatively inexpensive to solve numerically and can
sometimes be studied analytically. These models represent populations of immune
cells and bacteria as continuous concentration fields, rather than discrete
entities.  The states of the agents, such as age or size, can be described with age-
or size-structured PDEs, thereby representing heterogeneity among cells. PDE
models of \emph{Mycobacterium tuberculosis} infection \cite{catala2020reaction},
  microabscess formation \cite{pigozzo2012computational}, and the
immune response to an antigen \cite{su2009mathematical} have been developed. The
computational cost of such models usually does not scale with the population
size $M$, making these models appropriate when $M$ is large. Additionally, such
PDE models are often deterministic, so a single realization for a given
parameter set is sufficient in principle. A central difficulty with PDE models
is that the parameters, and sometimes even the form of the equations, are not
obviously determined by the underlying cell-level rules.

This difficulty can be addressed with systematic \emph{coarse-graining} or
upscaling of agent-based models into PDE descriptions, which effectively connect
the parameters between the two models by explicit formulae. There is a
significant body of literature on this subject in various contexts. Examples 
include oncolytic virotherapy \cite{morselli2023agent, morselli2026hybrid},
chemotaxis \cite{erban2004individual}, mechanical interactions between cells
\cite{lorenzi2020individual, chaplain2020bridging}, biological active matter
\cite{ridgway2023motility, zhao2026dynamic}, chemical signalling
\cite{dalwadi2020systematic},  dilute active suspensions
\cite{saintillan2014theory, fung2025foundation, yeo2026shear}, and the quantification of effective coefficients and emergent properties  \cite{murray2012classify, panigrahi2025intermittent, bruna2012excluded}.  However,
many of the rules which frequently occur in ABMs of bacterial infections have
not been connected with PDE descriptions. Therefore, we currently lack the
ability to pose accurate PDE models that correspond to the many ABMs of
bacterial infection, making it difficult to study large systems.

Here, we address this gap by constructing an ABM from several frequently occurring rules in the literature and systematically coarse-graining it. This analysis results in a reaction--diffusion PDE model that
reproduces the ensemble-averaged behaviour of the ABM in a biologically motivated parameter regime. We give explicit formulae for the coefficients of this PDE model
in terms of the original parameters of the ABM. We then perform another level of
coarse-graining in a certain continuum limit in the state space of the agents to
arrive at a novel continuum state-structured PDE. We emphasize that the goal of
this work is not to develop a new model for bacterial infections, but rather to
demonstrate how elements from existing ABMs can be systematically coarse-grained
into PDEs which are computationally more tractable than the original ABM.

The rest of this paper is organized as follows. Section~\ref{sec:abm}
defines the agent-based model and specifies the cell-level rules to be
coarse-grained. Section~\ref{sec:semidiscrete} derives the semidiscrete
reaction--diffusion system and computes the state-dependent ingestion rates from
an auxiliary diffusion problem. Section~\ref{sec:continuum} takes a continuum
limit in the internal variable to obtain a state-structured
reaction--diffusion system. Section~\ref{sec:examples} compares the ABM,
semidiscrete model, and continuum model in several numerical examples.
Section~\ref{sec:discussion} discusses limitations, extensions, and implications
for using coarse-grained PDEs to support ABM-based parameter studies.
Appendix~\ref{app:leading_order_insufficient} gives a steady-state example
showing that endpoint corrections can be required to preserve the qualitative
structure of the semidiscrete model. Appendix~\ref{app:kj_regular} gives  additional details on our analysis of the auxiliary diffusion problem in a special case.

\section{Agent-Based Model} \label{sec:abm}
We begin by constructing an ABM that consists of agents and rules governing the
agent behaviours. To illustrate the applicability of our coarse-graining
techniques, we build our ABM from rules that are used frequently in disease
models. We give a brief overview of these rules before getting into the
details. The agents in the system consist of immune cells  and extracellular
bacteria, which are confined to a domain $\Omega\subset\mathbb{R}^3$. The immune cell agents may represent, for example, macrophages, neutrophils, or T cells; the bacteria agents represent a pathogen, such as \mtb.   Each immune cell agent can ingest 
and clear bacteria. Additionally, each immune cell agent possesses an internal state
$I_j$, where the integer $j$ counts the number of ingested bacteria, and satisfies $0\leq j\leq N$. An immune cell in state $I_j$ with $j>0$ is said to be infected,
whereas an immune cell in state $I_0$ is non-infected. The internal state of an
immune cell agent can change over time as bacteria are cleared or ingested as we describe below. The immune cell agents also undergo natural death and are
introduced into the system with a certain probability per unit time, representing
influx \emph{via}, e.g., blood vessels.

Broadly, ingestion of bacteria by immune cells is accomplished through a process called  phagocytosis \cite{uribe2020phagocytosis}. This process is commonly represented in  ABMs by assuming that each immune cell agent ingests  some (possibly random) amount of bacteria  whenever there are bacteria agents nearby \cite{segovia2004identifying, petrucciani2024silico, cilfone2013multi, bowness2018modelling, ray2009synergy, alsassa2020modeling}. Following this convention, we assume that an immune cell agent in state $I_j$, with $j<N$,  ingests a bacterium agent when the two agents are within  an interaction distance $\bar{\rho}>0$. Agents in state $I_N$ cannot ingest additional bacteria.  This can be represented symbolically as
\begin{equation}
	I_j + B \overset{}{ \to } I_{j+1}, \quad j=0,\ldots, N-1. 
	\label{eqn:ABM_rule_ingestion}
\end{equation}
After ingestion, the extracellular bacterium agent is removed from the system
and the  state variable $j$ of the immune cell agent is incremented by one. Intracellular
bacteria are represented by the internal state $I_j$ of the immune cell agents,
rather than individual agents themselves.  

Following ingestion of bacteria, immune cells are capable of killing intracellular bacteria \cite{uribe2020phagocytosis}. We model this clearance of bacteria by equipping each
immune cell agent with a probability per unit time $q_j> 0$ of decreasing its
state variable by one. Some bacteria, such as \mtb, have the ability to survive
and replicate within phagosomes of macrophages \cite{schlesinger1993macrophage,
  kang2005human}. To model this, we equip each immune cell agent with another
probability per unit time $r_j$ of increasing its state variable by one. The
internal dynamics for each immune cell can be summarised symbolically by
\begin{subequations}
\begin{align}
	I_j&\overset{q_j}{\to} I_{j-1},  \quad j=1,\ldots N, \\
	I_j&\overset{r_j}{\to} I_{j+1}, \quad j=0,\ldots N-1.
\end{align}
\label{eqn:ABM_rule_internal}%
\end{subequations} 
We note that the index $j$ runs from $1$ to $N$ for $q_j$ because non-infected
immune cells have no bacteria to clear. Similarly, an immune cell cannot hold
more than $N$ bacteria, so $j$ runs from $0$ to $N-1$ for $r_j$. We impose $r_j>0$ for $j>0$, but allow  $r_0\geq0$ so that biologically realistic functional forms (such as exponential and logistic growth \cite{segovia2004identifying, petrucciani2024silico}) can be included in our analysis.    Rules similar to \eqref{eqn:ABM_rule_internal} are used in \cite{segovia2004identifying, petrucciani2024silico,cilfone2013multi, ray2009synergy, dancik2010parameter, shi2016agent}. 

Immune cell agents can be removed and added to the system. All immune cells undergo death with a probability per unit time $\beta_j$ that we allow to depend on the state $I_j$.  New immune cell agents are introduced into the system with probability per unit time, per unit volume, $a$. We assume that these new agents are not infected  and therefore initialize each new agent in the state $I_0$. Mathematically, these rules can be summarised by
\begin{subequations}
\begin{align}
	I_j&\overset{\beta_j}{\to} \emptyset, \label{eqn:ABM_rule_death}\\
	\emptyset& \overset{a}{\to} I_0, \label{eqn:ABM_rule_source}
\end{align}
\label{eqn:ABM_rule_death_source}%
\end{subequations}
where \(\emptyset\) denotes states outside the explicitly modelled system,
such as dead cells or cells outside the spatial domain. We remark that these rules, and
our subsequent analysis, can be straightforwardly generalized to include
situations where new cells are initialized with a random state $j$ according to
a prescribed probability function. In other ABMs, new immune cell agents are commonly introduced into the domain in specific source regions \cite{segovia2004identifying, bowness2018modelling, ray2009synergy}, representing e.g., blood vessels, which would be equivalent to allowing $a$ to depend on position $\mbf{x}$ in \eqref{eqn:ABM_rule_source}. Rules for cell death based on a probability per unit time, as in Eq.~\eqref{eqn:ABM_rule_death},  are also used in  \cite{carruthers2020stochastic, gillard2014modeling}. 

The positions of the immune cells and bacteria within the domain $\Omega$ at time $t$ are denoted by
$\mbf{x}^{(b)}_k(t)$ and $\mbf{x}^{(I)}_k(t)$, where $k$ indexes the agents.  We assume that both types of agents undergo pure Brownian
motion, which is common in the absence of chemical signalling  \cite{segovia2004identifying, petrucciani2024silico, cilfone2013multi, bowness2018modelling, ray2009synergy, dancik2010parameter}. The equations of motion for the position of each agent are therefore given by 
\begin{subequations}
\begin{align}
 	\mathrm{d}\mbf{x}_k^{(I)} &= \sqrt{2D_{j}}\mathrm{d}\mbf{W}^{(I)}_k(t), \quad k=1,\ldots, M_I, \label{eqn:ABM_rules_position_icell}\\
 	\mathrm{d}\mbf{x}_k^{(b)} &= \sqrt{2D_{b}}\mathrm{d}\mbf{W}^{(b)}_k(t), \quad k=1,\ldots, M_b,  
\end{align}
\label{eqn:ABM_rules_position}%
\end{subequations}
where $D_{b}$ and $D_{j}$ are the associated diffusion coefficients, $\mbf{W}_k^{(I)}(t)$ and $\mbf{W}_k^{(b)}(t)$ denote independent Wiener processes in $\mathbb{R}^3$, and  $M_I$ and $M_b$ are the (time-dependent) number of immune cell agents and bacteria agents, respectively.  The
diffusion coefficient for the immune cell agents is allowed to depend on the
state variable $j$ for generality because some ABMs in the literature assume that infected
immune cells move more slowly than non-infected immune cells \cite{segovia2004identifying, petrucciani2024silico, hoerter2025timing}, in line with experimental observations \cite{davis2009role}. For simplicity we assume that the agents are confined to the cube $\Omega:=[0,1]^3$, and we therefore impose
reflective boundary conditions. The position of an agent obeys
\eqref{eqn:ABM_rules_position} until it is removed from the system by either
ingestion, if the agent is a bacterium, or natural death, if the agent is an
immune cell. When a new agent is introduced into the system through
\eqref{eqn:ABM_rule_source}, its position is selected from a uniform
distribution in $\Omega$, consistent with constant source rate $a$ in Eq.~\eqref{eqn:ABM_rule_source}.

The complete ABM consists of the immune cell and bacteria agents as well as the
governing rules in
Eqs.~\eqref{eqn:ABM_rule_ingestion}--\eqref{eqn:ABM_rules_position}. To
facilitate our analysis in the following sections, we pose the model directly in
dimensionless form. In a typical ABM simulation of an {\mtb } infection, the
simulation domain has a side length on the order $1\, \mathrm{mm}$ with a time
horizon on the order of days. We therefore choose our characteristic length and
time scales to be $1\, \mathrm{mm}$ and $1000\, \mathrm{min}$, respectively.

\section{Semidiscrete Reaction--Diffusion System} \label{sec:semidiscrete}

Our goal in this section is to obtain a coarse-grained description of the ABM that is
capable of reproducing the average behaviour of the system. To this end, we
introduce the concentration of immune cells with internal state $j$ as
$u_j(\mbf{x},t)$, $j=0,\ldots, N$ and the concentration of bacteria as
$b(\mbf{x},t)$. The states of the immune cells remain discrete, but the number
and position of agents are represented by continuous quantities. We therefore
refer to this model as the \emph{semidiscrete} model.

The rules \eqref{eqn:ABM_rule_ingestion}--\eqref{eqn:ABM_rules_position} are essentially a molecular description of a reaction-diffusion network. We therefore pose a PDE reaction--diffusion system for the concentration of each type of agent. We expect this to be a good approximation of the ABM when the number of agents is sufficiently large, but in a dilute volume fraction regime where $\bar{\rho} \ll 1$. The reaction-diffusion system reads

\begin{subequations}
\begin{align}
	\pfo{\mbf{u}}{t} &= \mathcal{D}\nabla^2 \mbf{u} + \mathcal{M}\mbf{u} + (\mathcal{K}\mbf{u})b + \mbf{a}, \quad \mbf{x}\in \Omega, \label{eqn:u_semidiscrete} \\
	\pfo{b}{t} &= D_b \nabla^2 b - \sum_{j=0}^{N-1} k_j u_j b, \quad \mbf{x} \in \Omega \label{eqn:b_semidiscrete}	
\end{align}
where $\mbf{u} = \left(u_0, \ldots, u_N\right)^T$. The matrices $\mathcal{K}$ and $\mathcal{M}$ are given by
\begin{align}
	\mathcal{K} &= \begin{pmatrix}
		-k_0		&	0		&	0		&	\ldots 	&	0	\\
		k_0		&	-k_1		&	0		&	\ldots	&	0	\\
		0		&	k_1		&	\ddots	&			&	\vdots	\\
		\vdots 	&			&			&	0		&	0	\\
		0		&	\ldots	&	k_{N-2}	&	-k_{N-1}	&	0	\\
		0		&	\ldots	&	0		&	k_{N-1}	&	0		
	\end{pmatrix}, \\
	\mathcal{M} &= \begin{pmatrix}
		-r_0 - \beta_0 	& 	q_1 			& 	0		&	\ldots 			&	0		\\
		r_0 		&	 -r_1 -q_1 - \beta_1	& 	q_2 		&	\ldots 			&	0 		\\
		0		&		r_1 		& \ddots		&					& 	\vdots	\\
		\vdots	&				&			&	q_{N-1}				&	0 		\\
		0		&	\ldots			& r_{N-2}	& -r_{N-1} - q_{N-1} - \beta_{N-1}	& q_N 		\\
		0		&	\ldots			& 0			&	r_{N-1}			& -q_N - \beta_N		
	\end{pmatrix},
	\label{eqn:KM_def}
\end{align}
where the coefficients $k_j$ are to be determined in the course of our analysis. The matrix $\mathcal{D}$ and vector $\mbf{a}$ are given by
\begin{equation}
	  \mathcal{D} = \mathrm{diag}\left(D_0, \ldots, D_N\right), \quad \mbf{a} = (a,0,\ldots, 0)^T. \label{eqn:D_a_def}
\end{equation}

The terms involving the matrix $\mathcal{D}$ in Eq.~\eqref{eqn:u_semidiscrete}, as well as the first term on the RHS of \eqref{eqn:b_semidiscrete} represent the respective Brownian motion  of immune cells and bacteria in \eqref{eqn:ABM_rules_position}.  The terms involving $\mathcal{M}$ represent the bacteria-independent state transition rates \eqref{eqn:ABM_rule_internal} as well as  cell death Eq.~\eqref{eqn:ABM_rule_death}. In the context of chemical reactions, these terms represent first order reactions. The terms involving the matrix $\mathcal{K}$ in Eq.~\eqref{eqn:u_semidiscrete}, as well as the terms in the sum in Eq.~\eqref{eqn:b_semidiscrete}, encode the  rule  governing ingestion of bacteria in Eq.~\eqref{eqn:ABM_rule_ingestion}. More specifically, the coefficient $k_j$ represents the rate at which immune cells in state $j$ ingest bacteria, analogous to the rate constant of a bimolecular  chemical reaction. Finally, the vector $\mbf{a}$ represents the source of non-infected immune cells into the system.

We impose  no-flux boundary conditions for the semidiscrete model, consistent with the reflective boundary conditions in the ABM. These boundary conditions read
\begin{align}
	D_b \nabla b \cdot \hat{\mbf{N}} = D_j \nabla u_j \cdot \hat{\mbf{N}}=0, \quad \mbf{x} \in \partial\Omega, \quad j=0,\ldots, N,
	\label{eqn:semidiscrete_BCs}
\end{align}
\label{eqn:semidiscrete_system}%
\end{subequations}
where $\hat{\mbf{N}}$ is a unit normal vector on $\partial\Omega$. We also impose an appropriate initial condition, which will not be relevant for our analysis in this section. We remark that models similar to the well-mixed version of  \eqref{eqn:semidiscrete_system} have been studied in the context of phagocytosis of quartz particles \cite{da2020modelling, antunes2024modelling, tran1995mathematical} and {\mtb } infections in \cite{gammack2004macrophage}. In those works, the ingestion rates $k_j$ are input parameters with a prescribed dependence on $j$, rather than derived from a detailed microscopic model.  Our analysis reveals that the $k_j$ are not identical in general and their values depend on $r_j$, $q_j$, $D_j$, $\beta_j$, and $N$. Even in the special case where all of the input parameters are independent of $j$, the $k_j$ are still not identical.

To be able to use the semidiscrete model \eqref{eqn:semidiscrete_system}, we must
obtain numerical values for the rate constants $k_j$. Here we employ a Smoluchowski-type approach in which an effective ingestion rate is calculated from the steady state of an auxiliary diffusion problem \cite{erban2020stochastic, zhou1996theory, doi1976stochastic}. The basic idea is to
consider a system consisting of one bacterium immersed in a concentration field
of immune cells and to calculate the flux of immune cells out of the system; the
result is then proportional to the desired rate constants. This approach assumes a dilute system where the effects of the individual reactions on each other are negligible. A detailed analysis of the limitations of the theory is beyond the scope of this work; here we justify the approach \emph{a posteriori} through comparisons with the ABM in a biologically relevant parameter regime. The approach is
known to give good results for the simple bimolecular reaction $A+A\to B$
(see e.g., \cite{erban2020stochastic}) where it has been justified more rigorously with field operator methods \cite{doi1976stochastic}.  

We consider a single bacterium in a concentration field of immune cells. From
the frame of reference of the bacterium, the diffusion coefficient of an immune
cell in state $j$ is
\begin{equation}
	\bar{D}_j = D_b + D_j.
	\label{eqn:D_bar_component_def}
\end{equation} 
Then assuming that the side length of the domain is much larger than the
interaction radius, i.e.,  $\bar{\rho} \ll 1$, and on sufficiently long time
scales the concentrations of immune cells can be approximated by a smooth
solution of the system
\begin{subequations}
\begin{align}
	\mbf{0} &= \bar{\mathcal{D}} \nabla^2\mbf{u} + \mathcal{M} \mbf{u} + \mbf{a}, \quad |\mbf{x}|> \bar{\rho}, \label{eqn:b_frame_uj}\\
	0 &= \bar{D}_N\nabla^2 u_N - \left(q_N + \beta_N\right) u_N, \quad |\mbf{x}| < \bar{\rho},  \label{eqn:b_frame_uN_in}
\end{align}
where the matrix $\bar{\mathcal{D}}$ is defined by 
\begin{equation}
	\bar{\mathcal{D}} = \mathrm{diag}\left(\bar{D}_0, \ldots, \bar{D}_N\right),
	\label{eqn:D_bar_def}
\end{equation}
along with the boundary conditions
\begin{align}
	u_j(\mbf{x})& \text{ finite as } |\mbf{x}| \to \infty, \quad j=0,\ldots, N, \label{eqn:b_frame_BC_far} \\
	u_j(\mbf{x})&=0, \quad |\mbf{x}| = \bar{\rho}, \quad j=0,\ldots N-1, \label{eqn:b_frame_BC_near} \\
	u_N(\mbf{x}) & \text{ finite as } |\mbf{x}| \to 0.  \label{eqn:b_frame_BC_regular}
\end{align}
\label{eqn:b_frame_system}%
\end{subequations}
The far field condition \eqref{eqn:b_frame_BC_far} imposes  undisturbed immune cell concentrations far from the bacterium. The Dirichlet conditions in
Eq.~\eqref{eqn:b_frame_BC_near}, together with the fact that $u_N$ is the only
field defined in $|\mbf{x}| < \bar{\rho}$, enforces that only immune cells with
a full bacterial load ($j=N$) may be found inside the interaction radius
$\bar{\rho}$.  We estimate the reaction
constants $k_j$ as the relative flux of $u_j$ into the sphere of radius
$\bar{\rho}$,
\begin{equation}
	k_j = \frac{4\pi \bar{\rho}^2 \bar{D}_j|\nabla u_j|_{|\mbf{x}| =\bar{\rho}}}{\displaystyle \lim_{|\mbf{x}|\to \infty} u_j(\mbf{x})},  \quad j=0,\ldots,N-1.
	\label{eqn:reaction_rate_def}
\end{equation}
We proceed by solving Eq.~\eqref{eqn:b_frame_system} and substituting the solution into Eq.~\eqref{eqn:reaction_rate_def}.

To solve Eq.~\eqref{eqn:b_frame_system}, we seek a spherically symmetric
solution, which we denote by $\mbf{u}(\mbf{x}) = \mbf{u}(\rho)$, where
$\rho= |\mbf{x}|$.  Next, we note that Eq.~\eqref{eqn:b_frame_uN_in} is
decoupled from Eq.~\eqref{eqn:b_frame_uj} and can therefore be solved
directly. The general solution of \eqref{eqn:b_frame_uN_in} is
\begin{equation}
	u_N(\rho) = \frac{C}{\rho}\sinh\left( \frac{\mu \rho}{\bar{\rho}}\right) + \frac{\tilde{C}}{\rho}\cosh\left(\frac{\mu \rho}{\bar{\rho}}\right), \quad \mu = \sqrt{\frac{q_N + \beta_N}{\bar{D}_N}}\bar{\rho}, \quad \rho<\bar{\rho}.
\end{equation}
To satisfy the boundary condition \eqref{eqn:b_frame_BC_regular} for $u_N$, we set $\tilde{C}=0$. We determine $C$ after considering the solution in $\rho>\bar{\rho}$. 

Next we seek a solution of \eqref{eqn:b_frame_uj} of the form $\mbf{u}(\mbf{\rho}) = \rho^{-1}\mbf{v}(\rho)$. Substituting this ansatz into \eqref{eqn:b_frame_uj} yields the system
\begin{equation}
	\dft{\mbf{v}}{\rho} = \bar{\mathcal{M}}\mbf{v} + \bar{\mbf{a}}\rho, \quad \rho > \bar{\rho}, 
	\label{eqn:b_frame_v_eqn}
\end{equation}
where 
\begin{equation}
	\bar{\mathcal{M}} = -\bar{\mathcal{D}}^{-1} \mathcal{M}, \quad \bar{\mbf{a}} = -\bar{\mathcal{D}}^{-1}\mbf{a} = \left(-\frac{a}{\bar{D}_0}, 0, \ldots, 0\right)^T. 
	\label{eqn:M_a_bar_def}	
\end{equation}
The form of the general solution of \eqref{eqn:b_frame_v_eqn} depends on the eigenvalues of the tridiagonal matrix $\bar{\mathcal{M}}$. We will show that these eigenvalues are real, positive, and distinct when $r_j>0$;  the special case  $r_0=0$ will be discussed at the end of this section. To accomplish this, we note that the products of the entries on the sub- and super-diagonals of $\bar{\mathcal{M}}$ are given by
\begin{equation}
	\bar{\mathcal{M}}_{j,j+1}\bar{\mathcal{M}}_{j+1,j}=\frac{q_j r_{j-1}}{\bar{D}_j \bar{D}_{j-1}}>0, \quad j=1,\ldots, N,
\end{equation}
where the entries $\bar{\mathcal{M}}_{i,j}$ are indexed from one.  Because these quantities are positive for all $j$, $\bar{\mathcal{M}}$ has distinct, real eigenvalues (see \cite{horn2012matrix}, Ch.~3.1). To show that the eigenvalues are positive, we note that the matrix $\mathcal{A}$ defined by the similarity transform 
\begin{equation}
	\mathcal{A} = P^{-1}\bar{\mathcal{M}} P, \quad P=\mathrm{diag}\left(p_0, \ldots, p_N\right), \quad p_0 =1, \quad \frac{p_{j+1}}{p_j}=\sqrt{\frac{\bar{D}_{j}r_j}{q_{j+1}\bar{D}_{j+1}}}, \label{eqn:A_similar}
\end{equation}
is tridiagonal and symmetric. We  show that $\mathcal{A}$ has positive eigenvalues by showing that  the  determinants of the leading principal minors of $\mathcal{A}$  are positive. By induction on $j$,  the determinant of the $j^\text{th}$ leading principal  minor of $\mathcal{A}$, denoted by $M_j$, is given recursively by 
\begin{subequations} 
\begin{align}
	M_{j+1} &= \frac{r_j + \beta_j}{\bar{D}_j}M_{j} + \frac{q_j}{\bar{D}_j} F_{j}, \quad j=1,\ldots, N-1, \\
	F_{j+1} &= \frac{\beta_j}{\bar{D}_j} M_{j} + \frac{q_j}{\bar{D}_j}F_{j},  \quad j=1,\ldots, N-1,
\end{align}
\label{eqn:minor_recursion}%
\end{subequations}
with initial conditions 
\begin{equation}
	M_1 = \frac{r_0+ \beta_0}{\bar{D}_0}, \quad F_1 = \frac{\beta_0}{\bar{D}_0}.
\end{equation}
Because $F_1$ and $M_1$ are positive, the recursion relations Eq.~\eqref{eqn:minor_recursion} imply that all $M_j$ are  positive. Moreover, the determinant of $\mathcal{A}$ is
\begin{equation}
	\det \mathcal{A} = \frac{\beta_N}{\bar{D}_N}M_{N} + \frac{q_N}{\bar{D}_N} F_{N}>0.
\end{equation}
Thus $\mathcal{A}$ is positive definite and hence has positive eigenvalues. As $\bar{\mathcal{M}}$ is similar to $\mathcal{A}$ by Eq.~\eqref{eqn:A_similar}, $\bar{\mathcal{M}}$ also has positive eigenvalues. Because the eigenvalues of $\bar{\mathcal{M}}$ are real, positive, and distinct, the general solution to \eqref{eqn:b_frame_v_eqn} can be written as
\begin{subequations}
\begin{equation}
	\mbf{v}(\rho) = \sum_{m=0}^N \mbf{e}_m \left(C_m e^{-\sqrt{\lambda_m}\rho} + \tilde{C}_m e^{\sqrt{\lambda_m}\rho}\right) -\left(\bar{\mathcal{M}}^{-1}\bar{\mbf{a}}\right) \rho, \quad \rho> \bar{\rho},
\end{equation}
\label{eqn:b_frame_v_soln}%
\end{subequations}
where $(\lambda_m, \mbf{e}_m)$ are the eigenpairs of $\bar{\mathcal{M}}$. The
coefficients $C_m$, and $\tilde{C}_m$ are constants to be determined.  To satisfy the far-field conditions \eqref{eqn:b_frame_BC_far}, we must set $\tilde{C}_m=0$ for all $m=0,\ldots, N$. Hence $\mbf{u} = \mbf{v}/\rho \to -\bar{\mathcal{M}}^{-1}\bar{\mbf{a}}$ as $\rho\to \infty$.

To determine the $N+2$ unknown constants  $C$ and $C_m$, we impose the Dirichlet boundary  conditions in Eq.~\eqref{eqn:b_frame_BC_near} as well as continuity of $u_N$ and its derivative across $\rho=\bar{\rho}$.  Imposing the latter yields
\begin{subequations}
\begin{align}
	&\sum_{m=0}^NC_m e_{Nm}e^{-\mu_m} - C\sinh\mu =  \bar{\rho}\left(\bar{\mathcal{M}}^{-1}\bar{\mbf{a}}\right)_N, \\
	&\sum_{m=0}^N  e_{Nm}(1 + \mu_m)C_me^{-\mu_m} + \left(\mu \cosh \mu - \sinh \mu\right)C  =0,
\end{align}
\label{eqn:b_frame_linear_system1}
\end{subequations}
where $e_{jm}$ is the $j^\text{th}$ component of $\mbf{e}_m$ and $\mu_m$ is defined by
\begin{equation}
	\mu_m = \sqrt{\lambda_m}\bar{\rho}.
\end{equation}
Next, we impose the Dirichlet conditions \eqref{eqn:b_frame_BC_near} at $\rho = \bar{\rho}$ on $u_0, \ldots u_{N-1}$. This yields 
\begin{equation}
	\sum_{m=0}^NC_m e_{jm}e^{-\mu_m} =  \bar{\rho}\left(\bar{\mathcal{M}}^{-1}\bar{\mbf{a}}\right)_j, \quad j=0, \ldots, N-1.
	\label{eqn:b_frame_linear_system2}
\end{equation}
Eqs.~\eqref{eqn:b_frame_linear_system1} and \eqref{eqn:b_frame_linear_system2}
constitute an $(N+2)$-by-$(N+2)$ linear system that can be solved
uniquely for $C$ and $C_m$. In principle, the solution of \eqref{eqn:b_frame_system} is now known.

We now estimate the effective ingestion rates $k_j$ using \eqref{eqn:reaction_rate_def}. Inserting the components of $\mbf{u}= \rho^{-1}\mbf{v}$, with $\mbf{v}$ given in \eqref{eqn:b_frame_v_soln}, we obtain
\begin{equation}
  k_j = -\frac{4\pi \bar{D}_j}{\left(\bar{\mathcal{M}}^{-1}\bar{\mbf{a}}\right)_j}\sum_{m=0}^Ne_{jm}C_m(1 + \mu_m)e^{-\mu_m} , \quad j=0,\ldots, N-1.
  \label{eqn:effective_reaction_rate}
\end{equation}
Eq.~\eqref{eqn:effective_reaction_rate} constitutes the main result of this
section. It shows that the effective ingestion rates are not free parameters of
the coarse-grained model; rather, they are determined by the microscopic
diffusion coefficients, intracellular transition rates, death rates,
 and interaction radius. We note the parameter $a$ actually drops out of Eq.~\eqref{eqn:effective_reaction_rate}. This is because the solution of Eq.~\eqref{eqn:b_frame_system} is proportional to $\mbf{a}=(a,0,\ldots,0)^T$ and the ingestion rates $k_j$ are independent of the overall scaling of $\mbf{u}$.

To obtain a numerical value for each $k_j$ from
Eq.~\eqref{eqn:effective_reaction_rate}, we first compute the eigenpairs
$(\mbf{e}_m,\lambda_m)$ of $\bar{\mathcal{M}}$, defined in
Eqs.~\eqref{eqn:M_a_bar_def}, \eqref{eqn:D_bar_def},
\eqref{eqn:D_bar_component_def}, and \eqref{eqn:KM_def}. We then solve the
linear system in Eqs.~\eqref{eqn:b_frame_linear_system1} and
\eqref{eqn:b_frame_linear_system2} for $C$ and $C_m$, $m=0,\ldots,N$, and
substitute these constants into Eq.~\eqref{eqn:effective_reaction_rate}. With
the values of $k_j$ fixed, the semidiscrete system
\eqref{eqn:semidiscrete_system} can be solved numerically.

Before moving on, we briefly discuss the special case $r_0=0$. In this case,
the matrix $\bar{\mathcal{M}}$ still has real, positive eigenvalues, but they may
not be distinct. The general solution of Eq.~\eqref{eqn:b_frame_v_eqn} can be
expressed in terms of the generalized eigenvectors of $\bar{\mathcal{M}}$. However, the form of the solution of \eqref{eqn:b_frame_v_eqn} can instead be obtained directly by inspection because $(1,0,\ldots,0)^T$ is now an eigenvector of $\bar{\mathcal{M}}$.  The solution for $\mbf{u}$ now reads
\begin{equation}
	\mbf{u} = (1,0,\ldots, 0)^T \frac{a}{\beta_0}\left(1 + C\frac{\bar{\rho}}{\rho} \exp\left(-\sqrt{\frac{\beta_0}{\bar{D}_0}}(\rho - \bar{\rho})\right)\right).
\end{equation}
The constant $C=-1$ is obtained by imposing the boundary conditions in
Eq.~\eqref{eqn:b_frame_BC_near}.
Here the effective ingestion rates cannot be
computed using \eqref{eqn:reaction_rate_def} because the denominator vanishes
for $j=1,\ldots,N-1$. However, we show in Appendix \ref{app:kj_regular} that $k_j=\mathcal{O}(1)$ as  $r_0\to0^+$.
We therefore use a small value of $r_0$ in the calculation of $k_j$ in
situations where $r_0=0$.

\section{Continuum Limit} \label{sec:continuum}

In this section, we consider a continuum limit of \eqref{eqn:semidiscrete_system} where the maximum intracellular bacterial load is large, i.e., $N\gg 1$. This is a reasonable assumption for many biological systems. For example, alveolar macrophages can ingest between 20--40 {\mtb } before bursting \cite{repasy2013intracellular}, and Kupffer cells  can ingest  $N>50$ \emph{Salmonella choleraesuis} bacteria \cite{nnalue1992salmonella}.

In the regime where $N\gg 1$, we identify the discrete variable $j=0,\ldots, N$ with a continuous version $s\in [0,1]$. More specifically, we set
\begin{equation}
	s\sim \frac{j}{N},
\end{equation}
where equality  holds only when $s$ takes on integer multiples of $1/N$. We also define functions $u$, $k$, $q$, $r$, and $\beta$ that extend the discrete versions $u_j$, $k_j$, $q_j$, $r_j$, and $\beta_j$ to non-integer values of $j$, i.e.,
\begin{equation}
	u(s,\mbf{x},t) \sim u_j(\mbf{x},t), \quad k(s) \sim k_j, \quad q(s)\sim q_j, \quad r(s)\sim r_j, \quad \beta(s) \sim \beta_j, \quad s\in [0,1].
	\label{eqn:continuum_coeffs}
\end{equation}
We can formulate Eq.~\eqref{eqn:u_semidiscrete} in terms of these continuous functions as
\begin{align}
	\pfo{u}{t} =& D(s)\nabla^2 u + \left[k(s-\Delta s)b + r(s-\Delta s)\right]u(s-\Delta s)  - \left[ k(s)b + r(s) + q(s)\right]u(s) \nonumber \\
	&+ q(s+\Delta s)u(s+\Delta s) - \beta(s)u(s), \quad s\in (\Delta s,1-\Delta s), \quad \mbf{x}\in \Omega,
	 \label{eqn:u_semidiscrete_interior_eqn}
\end{align}
where $\Delta s = N^{-1}$ and we suppress the dependence of $u$ on $t$ and $\mbf{x}$ for notational brevity. The domain of Eq.~\eqref{eqn:u_semidiscrete_interior_eqn} with respect to the $s$--coordinate excludes regions within $\Delta s$ of the  boundaries at $s=0$ and $s=1$ to avoid evaluating $u$, $k$, $q$, and $r$ outside of their domains. Thus the equations for $u_0$ and $u_N$ are effectively excluded in \eqref{eqn:u_semidiscrete_interior_eqn}.

Next, we take a continuum limit of \eqref{eqn:u_semidiscrete_interior_eqn} in
the regime $N\gg 1$, or equivalently, $\Delta s\to 0^+$. We assume that the
functions $u$, $k$, $q$, $r$, and $\beta$ are sufficiently smooth to be expanded in a
Taylor series. We also assume that these functions do not vary rapidly with $s$,
in particular their derivatives should be $\mathcal{O}(1)$ as $N\to
\infty$. Equivalently, the discrete versions of these quantities,  $u_j$, $k_j$, $q_j$, $r_j$, and $\beta_j$, should vary slowly with $j$, i.e., $|q_{j+1}-q_j| = \mathcal{O}(N^{-1})$, and similarly for the other quantities. This restriction is straightforward to satisfy for the input parameters $q_j$, $r_j$, and $\beta_j$. However, both $u_j$ and $k_j$ can have boundary layers as we will discuss in Section \ref{sec:examples}.  For now, we proceed under the assumption that the derivatives of $u$, $k$, $q$, $r$, and $\beta$ are $\mathcal{O}(1)$ as $N\to \infty$.

Taylor expanding each of the non-local terms in \eqref{eqn:u_semidiscrete_interior_eqn} about $\Delta s=0$ and neglecting $\mathcal{O}(N^{-3})$ terms  yields
\begin{equation}
	\pfo{u}{t} = D(s)\nabla^2 u + \frac{1}{2N^2}\pft{}{s}\left(f(s,b)u\right) - \frac{1}{N}\pfo{}{s}\left(g(s,b)u\right) - \beta(s)u, \quad  (s,\mbf{x})\in (0,1)\times \Omega,
	\label{eqn:u_continuum}
\end{equation}
where the functions $f$ and $g$ are defined as
\begin{equation}
	f(s,b) = q(s) + r(s) + k(s)b, \quad g(s,b) = -q(s) + r(s) +k(s)b. \label{eqn:fg_def}
\end{equation}
Eq.~\eqref{eqn:u_continuum} requires two additional boundary conditions on the state boundaries $s=0$ and $s=1$. To find the appropriate conditions, we demand that the mass in the continuum system evolves consistently with the discrete system. In particular, we observe that summing the equations in Eq.~\eqref{eqn:u_semidiscrete} over $j$ yields
\begin{equation}
	\frac{1}{N}\sum_{j=0}^N \pfo{u_j}{t} = \frac{1}{N}\sum_{j=0}^N D_j\nabla^2 u_j  - \frac{1}{N}\sum_{j=0}^N\beta_j u_j + \frac{1}{N}a.
	\label{eqn:semidiscrete_mass}
\end{equation}
If we integrate \eqref{eqn:u_continuum} in $s$ over $(0,1)$, we obtain
\begin{equation}
	\int_0^1 \pfo{u}{t}\,\mathrm{d}s = \int_0^1 D(s)\nabla^2 u \,\mathrm{d}s  - \int_0^1 \beta(s) u\,\mathrm{d}s +  \left[\frac{1}{2N^2}\pfo{}{s}\left( f(s,b) u\right) - \frac{1}{N}g(s,b) u \right]\bigg|_0^1.
	\label{eqn:continuum_mass}
\end{equation}
Eq.~\eqref{eqn:semidiscrete_mass} is a discrete approximation of the integral terms in \eqref{eqn:continuum_mass}. Therefore, one way to demand consistency between the two is to set 
\begin{subequations}
\begin{align}
	\frac{1}{2N^2}\pfo{}{s}\left( f(s,b) u\right) - \frac{1}{N}g(s,b) u &=-\frac{a}{N}, \quad s=0, \label{eqn:continuum_BC_s0}\\
	\frac{1}{2N^2}\pfo{}{s}\left( f(s,b) u\right) - \frac{1}{N}g(s,b) u &= 0, \quad s=1.
	\label{eqn:continuum_BC_s1}
\end{align}
\label{eqn:continuum_BC_s}%
\end{subequations}
Physically, the boundary condition \eqref{eqn:continuum_BC_s0} represents an influx of non-infected immune cells through the state boundary $s=0$ \emph{via} the source rule, given in \eqref{eqn:ABM_rule_source}. Eq.~\eqref{eqn:continuum_BC_s1} is a no-flux condition on the $s=1$ state boundary.

Next, we reformulate Eq.~\eqref{eqn:b_semidiscrete} for $b$ in terms of the continuous variables $k$ and $u$. To this end, we rewrite the sum as
\begin{equation}
	\pfo{b}{t} = D_b \nabla^2 b - \sum_{j=0}^{N-1} k(s_j) u(s_j) b,
	\label{eqn:b_eqn_rewrite}
\end{equation}
where $s_j = j/N$ and we again suppress  the dependence of $u$ and $b$ on $\mbf{x}$ and $t$. Intuitively, the sum in Eq.~\eqref{eqn:b_eqn_rewrite} is simply the leading order approximation of the integral $\int_0^1 k(s)u(s)\,\mathrm{d}s$. However, simply replacing the sum in Eq.~\eqref{eqn:b_eqn_rewrite} with an integral assumes that the functions $k$ and $u$ do not vary rapidly as $N\to \infty$. To see this explicitly, we proceed by considering the integral $I$ defined by 
\begin{equation}
	I:=\int_0^1 h(s)\,\mathrm{d}s,
\end{equation}
where $h(s) $ is a smooth function whose values and derivatives are $\mathcal{O}(1)$ as $N\to \infty$. We can decompose the integration into $N$ subintervals of width $N^{-1}$ as
\begin{equation}
	I = \sum_{j=0}^{N-1} I_j, \quad I_j:=\int_{s_j}^{s_{j+1}} h(s)\,\mathrm{d}s. 
\end{equation}
Assuming that $N\gg 1$, we expand $h$ in a Taylor series about $s=s_j$  and evaluate the integral over each subinterval separately by integrating term-by-term:
\begin{equation}
	I_j = \int_{s_j}^{s_{j+1}} \sum_{n=0}^\infty \frac{h^{(n)}(s_j)}{n!}(s- s_j)^n = \sum_{n=0}^\infty \frac{h^{(n)}(s_j)}{(n+1)!}N^{-(n+1)},\quad j=0,\ldots, N-1.
\end{equation}
Then summing over $j=0,\ldots, N-1$ and extracting the leading order terms, we find
\begin{equation}
	I = \sum_{j=0}^{N-1} \sum_{n=0}^\infty \frac{h^{(n)}(s_j)}{(n+1)!}N^{-(n+1)} = \frac{1}{N}\sum_{j=0}^{N-1} h(s_j) + \sum_{j=0}^{N-1} \sum_{n=1}^\infty \frac{h^{(n)}(s_j)}{(n+1)!}N^{-(n+1)}.
	\label{eqn:I_sum}
\end{equation}
The first sum on the RHS is $\mathcal{O}(1)$ because $h(s)=\mathcal{O}(1)$. Rearranging \eqref{eqn:I_sum}, we obtain
\begin{equation}
	\frac{1}{N}\sum_{j=0}^{N-1} h(s_j)  = \int_0^1 h(s)\,\mathrm{d}s - \frac{1}{2N^2}\sum_{j=0}^{N-1} h'(s_j)  + \mathcal{O}(N^{-2}),
\end{equation}
where again $h(s)=\mathcal{O}(1)$ implies that second term is $\mathcal{O}(N^{-1})$ and the neglected terms are $\mathcal{O}(N^{-2})$ as long as the derivatives of $h$ are $\mathcal{O}(1)$. Finally, because $h$ is smooth, we can apply this formula to $h'$ to approximate the sum over $h'(s_j)$, which leads to 
\begin{equation}
	\frac{1}{N}\sum_{j=0}^{N-1} h(s_j)  =  \int_0^1 h(s)\,\mathrm{d}s - \frac{h(1) - h(0)}{2N}+ \mathcal{O}(N^{-2}).
	\label{eqn:integral_approx}
\end{equation}
Thus we see that replacing the sum with an integral neglects the $\mathcal{O}(N^{-1})$ term on the RHS. We retain only the leading order term in our subsequent analysis.  In general, we will see later that the system can form boundary layers in the  $s$--coordinate. The presence of these boundary layers violates the assumption that the derivatives of $h(s)$  are $\mathcal{O}(1)$, which implies that retaining terms beyond leading order in Eq.~\eqref{eqn:integral_approx} will not yield higher order convergence as one might naively expect. We show in Appendix \ref{app:leading_order_insufficient} that there is a special case where the $\mathcal{O}(N^{-1})$ term is required for $\mathcal{O}(1)$ agreement with the semidiscrete system, even when the system develops a boundary layer.

To obtain an equation for $b$ in terms of $u$, we apply Eq.~\eqref{eqn:integral_approx} to the sum in \eqref{eqn:b_eqn_rewrite} and retain only the leading order terms. Our full  system then reads
\begin{subequations}
\begin{align}
	\pfo{u}{t} &= D(s)\nabla^2 u + \frac{1}{2N^2}\pft{}{s}\left(f(s,b)u\right) - \frac{1}{N}\pfo{}{s}\left(g(s,b)u\right) - \beta(s) u,\quad  (s,\mbf{x})\in (0,1)\times \Omega,\label{eqn:continuum_system_u}\\
	\pfo{b}{t} &= D_b \nabla^2 b - b N\int_0^1 k(s)u(s,\mbf{x},t)\,\mathrm{d} s, \quad \mbf{x} \in \Omega, \label{eqn:continuum_system_b}
\end{align}
with no-flux boundary conditions on the physical boundaries, given by
\begin{equation}
	D(s)\nabla u\cdot \hat{\mbf{N}}  = D_b\nabla b \cdot \hat{\mbf{N}} = 0, \quad \mbf{x} \in \partial \Omega,
	\label{eqn:continuum_BC_space}		
\end{equation}
as well as Robin boundary conditions for the state-space boundaries, given by
\begin{align}
	\frac{1}{2N}\pfo{}{s}\left( f(s,b) u\right) - g(s,b) u &= - a, \quad s=0, \label{eqn:continuum_BC_s0_final}\\
	\frac{1}{2N}\pfo{}{s}\left( f(s,b) u\right) - g(s,b) u &=0, \quad s=1.
	\label{eqn:continuum_BC_s1_final}
\end{align}
\label{eqn:continuum_system}
\end{subequations}

The PDE system Eq.~\eqref{eqn:continuum_system} is the main result of this section. We refer to this system as the \emph{continuum} system because it represents the cell and bacteria populations as fully continuous quantities in both the physical space and the state space. The continuum system is formally valid in the regime $N\gg 1$ and when the quantities indexed by $j$ in the semidiscrete system \eqref{eqn:semidiscrete_system} do not vary rapidly. 

\section{Examples} \label{sec:examples}

To validate the semidiscrete and continuum PDE systems against the ABM, we compare the predictions of each of the three models. We give three examples in this section. In the first example, we compare the semidiscrete system with the ABM using a biologically relevant parameter set to validate our calculation of the effective ingestion rates. In our second example, we compare the semidiscrete and continuum PDE systems for a generic parameter set that satisfies our smoothness assumptions in the $N\gg 1$ limit. We show that the solutions develop boundary layers, causing localized disagreement between the two models. In the final example, we compare all three models in a biologically relevant parameter regime to demonstrate that both the semidiscrete and continuum models are capable of reproducing the ensemble averaged behaviour of the ABM.

To compare the  dynamics of the three models in the following examples, we define several variables to track throughout the simulation. First, we define the state-structured cell counts $U(s,t)$ and $U_j(t)$ as
\begin{equation}
	U(s,t) = \int_\Omega u(s,\mbf{x},t)\,\mathrm{d}\mbf{x}, \quad U_j(t) = \int_\Omega u_j(x,t)\,\mathrm{d}\mbf{x}, \label{eqn:u_counts_def}
\end{equation}
as well as the total cell and  bacteria counts
\begin{equation}
	M^{(\mathrm{C})}(t) = N\int_0^1 U(s,t)\,\mathrm{d}s, \quad M^{(\mathrm{SD})}(t) = \sum_{j=0}^N U_j(t), \quad B(t) = \int_\Omega b(\mbf{x},t)\,\mathrm{d}\mbf{x},
	\label{eqn:total_counts}
\end{equation}
where superscripts `$\mathrm{C}$' and `$\mathrm{SD}$' denote ``continuum" and ``semidiscrete," respectively. The cell and bacteria counts for the ABM are obtained simply by tracking the total number of agents in each state and ensemble averaging over many independent realizations. The cell and bacteria densities for the ABM are estimated by binning  and ensemble averaging. We use an ensemble size of 200 for all ABM simulations. 

For simplicity, we consider examples that are effectively one dimensional in space for both the continuum and semidiscrete  systems.  We emphasize
that the full three spatial dimensions must be used in the ABM for the
derivation of the ingestion rates in \eqref{eqn:effective_reaction_rate} to be
valid. The ABM is implemented in PhysiCell \cite{physicell}. Both the  semidiscrete and continuum systems are 
discretized in space with a Galerkin finite element method with quadratic Lagrange elements. Time
integration for the semidiscrete system is performed with Strang splitting, using SDIRK2 for the diffusion
subproblems and a second-order explicit Heun method for the reaction
subproblems. Time integration for the continuum system is performed monolithically using a second-order explicit Heun method. The implementation uses the \verb|tostii| operator splitting
library for \verb|dealii| \cite{moayeri2025tost, arndt2022deal}.

\subsection{Example 1: Comparison between the ABM and the Semidiscrete System}

\begin{figure}
\centering
\includegraphics[width=0.6\textwidth]{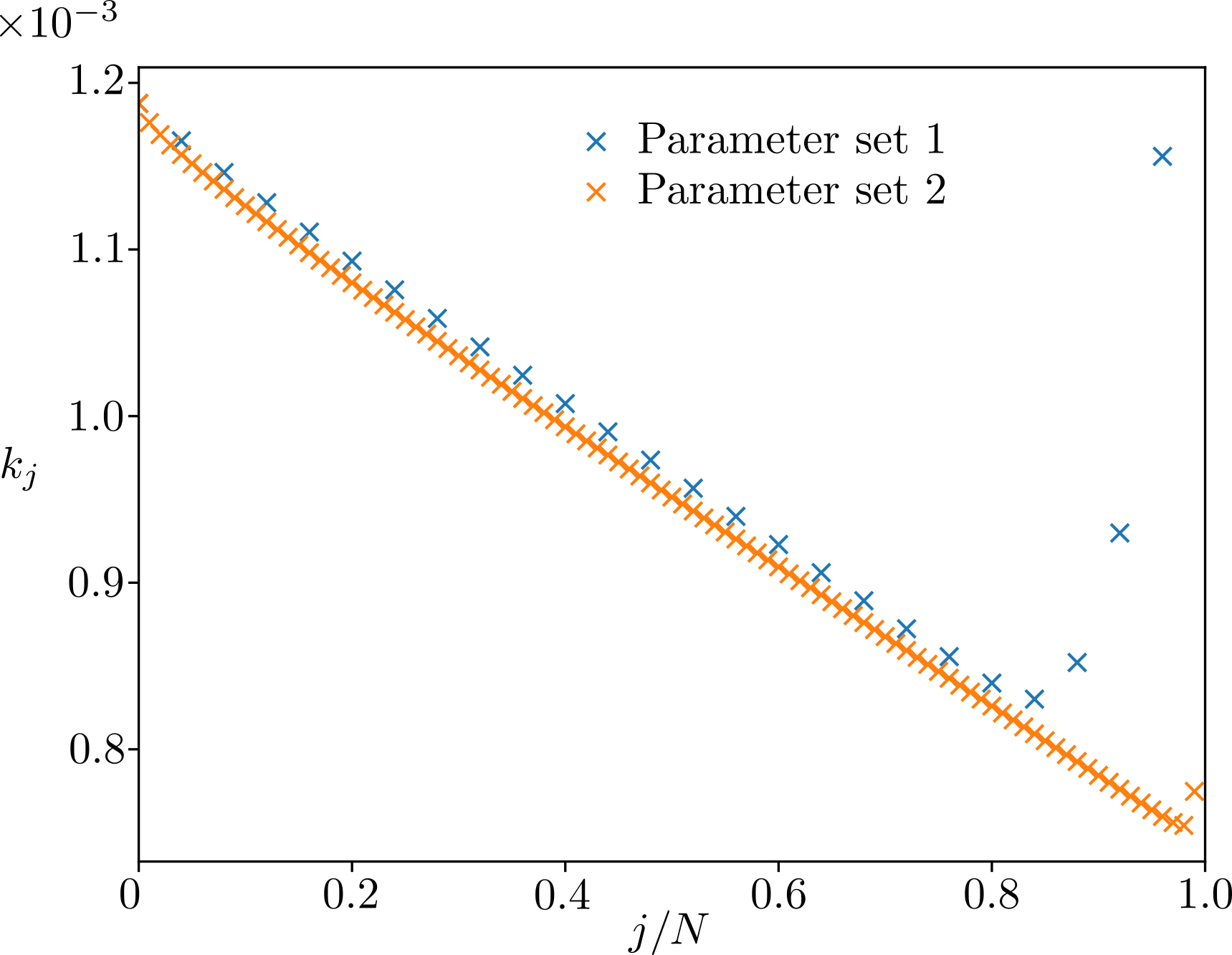}
\caption{Effective ingestion rate coefficients calculated from \eqref{eqn:effective_reaction_rate} using parameter set 1 in Eq.~\eqref{eqn:eg1_params} and parameter set 2 in  Eq.~\eqref{eqn:eg3_params}.}
\label{fig:eg1_reaction_rates}
\end{figure}

We first compare the results from the ABM and semidiscrete system
\eqref{eqn:semidiscrete_system}. We use the following parameter set, which is
motivated from the ABMs of {\mtb } infection in \cite{segovia2004identifying,
  petrucciani2024silico}:
\begin{align}
	q_j &= 2, \quad r_j = j\left(1 - \frac{j}{N+30}\right), \quad \beta_j = 5\times 10^{-2},  \nonumber \\
	D_j &= 5\times 10^{-3}\left(1- \frac{j}{2N}\right), \quad D_b = 2\times10^{-3}, \quad a = 10, \quad N=25.
	\label{eqn:eg1_params}
\end{align}
We recall that time and space are scaled by our characteristic  time and length scales of $1000\, \mathrm{min}$ and $1\,\mathrm{mm}$, respectively. The form of the intracellular replication rate of bacteria $r_j$ is taken from
\cite{segovia2004identifying}. The clearance
rate $q_j$ of intracellular bacteria depends on the specific type of immune cell
as well as its activation status; here we select an intermediate value based on
the values given in \cite{segovia2004identifying, petrucciani2024silico}. The
movement speed of immune cells decreases as their intracellular load increases
\cite{davis2009role}. We therefore impose a simple linear form for $D_j$. The overall scaling
for \(D_b\) and \(D_j\) is chosen to represent relatively slow undirected motion
in the absence of chemical signalling. Typical values for the lifespan of an
alveolar macrophage are between 10 and 100 days \cite{segovia2004identifying,
  petrucciani2024silico}. We obtain an order of magnitude approximation of the
death rate as the reciprocal of the lifespan.  The ratio of source and
death rates  determines the approximate number of immune cells in the system; we select the parameter $a$
such that there are on the order of 100 immune cells. The interaction radius
$\bar{\rho}\approx 0.013$ is chosen to be $1.25$ times the sum of the radii of an immune cell
and a bacterium. For immune cells we assume a radius of $10\,\mu\mathrm{m}$  and for the bacteria we calculate the radius by assuming a spherical volume of $0.35\,\mu\mathrm{m}^3$; these values are based on the values reported in \cite{petrucciani2024silico, segovia2004identifying} and the references therein. The effective ingestion rates $k_j$ are calculated from
\eqref{eqn:effective_reaction_rate} and shown in Figure
\ref{fig:eg1_reaction_rates}. We note that this example falls into the special case where $r_0=0$. Following our discussion at the end of Section \ref{sec:semidiscrete}, we set $r_0 \approx 10^{-6}$ in the calculation of $k_j$, and set $r_0=0$ in the numerical solution of the semidiscrete system.

For the semidiscrete system \eqref{eqn:semidiscrete_system}, we choose the initial condition 
\begin{subequations}
\begin{align}
	u_j(\mbf{x},0) &= C_1 U_{j}(0)\exp\left(-50\left(x- \frac{3}{4}\right)^2\right), \label{eqn:eg1_uj_IC}\\
	b(\mbf{x},0) &= C_2 B(0)\exp\left(-50\left(x- \frac{1}{4}\right)^2\right), \label{eqn:eg1_b_IC}
\end{align}
where $\mbf{x}=(x,y,z)$. The quantities  $B(0)$ and $U_j(0)$ are given by
\begin{equation}
	B(0)=5000, \quad 	U_{j}(0) = \begin{cases}
		1, \quad j=1,8,\\
		5, \quad j=2,7, \\
		10, \quad j=3,6, \\
		34, \quad j=4,5, \\
		0, \quad \mathrm{otherwise},
	\end{cases}\label{eqn:eg1_IC_u}
\end{equation}
and the normalization constants $C_1$ and $C_2$ are chosen  such that $\int_\Omega u_j(\mbf{x},0)\,\mathrm{d}x = U_{j}(0)$ and $\int_\Omega b(\mbf{x},0)\,\mathrm{d}x = B(0)$. We note that $C_1=C_2$ by symmetry.
\label{eqn:eg1_IC}
\end{subequations}

For the ABM, we sample agent positions from distributions consistent with
the  initial condition \eqref{eqn:eg1_IC}.
The system is initialized with $B(0)=5000$ bacteria, and
$U_j(0)$ immune cells with internal state $I_j$. The initial $x$-coordinates of each
agent are selected from a suitably normalized version of \eqref{eqn:eg1_uj_IC}
and \eqref{eqn:eg1_b_IC}, while the $y$ and $z$ coordinates are selected from
uniform distributions.

\begin{figure}[h!]
\centering
\includegraphics[width=0.95\textwidth]{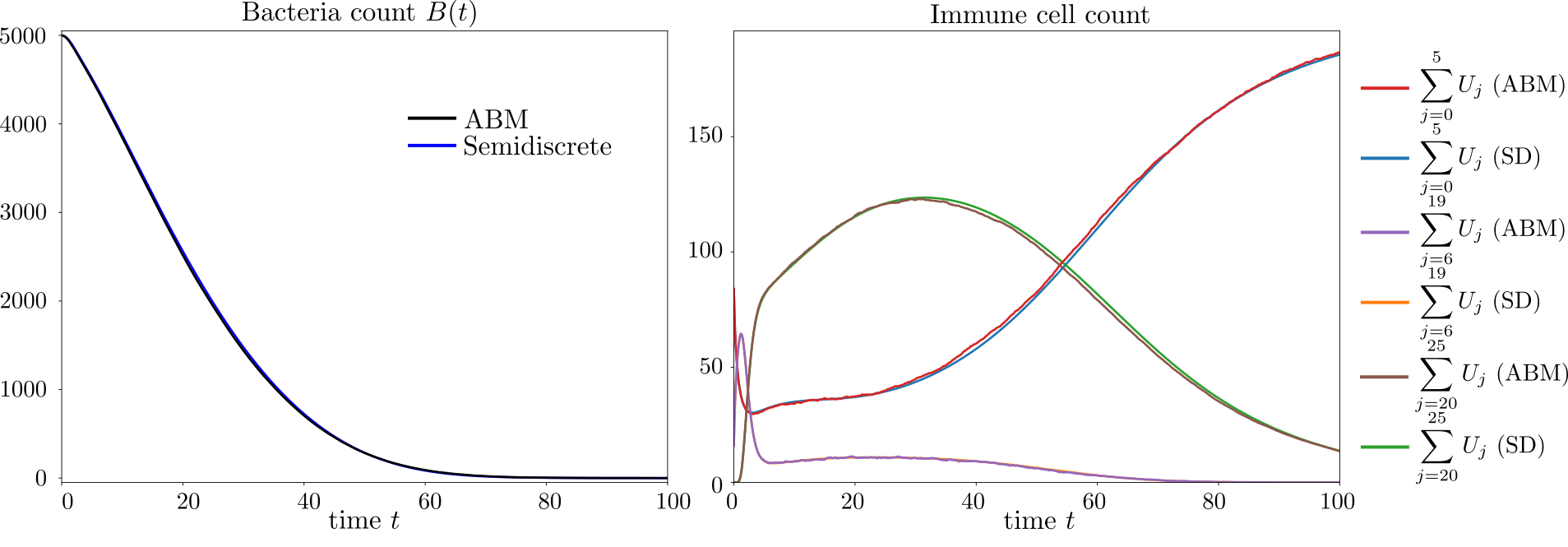}
\caption{Comparison of cell and bacteria counts (left and right panels respectively) over time showing excellent agreement between ABM and the semidiscrete (SD) system. The state structured cell count $U_j$ is defined in Eq.~\eqref{eqn:u_counts_def}.  }
\label{fig:eg1_ub}
\end{figure}

\begin{figure}[h!]
\centering
\includegraphics[width=0.95\textwidth]{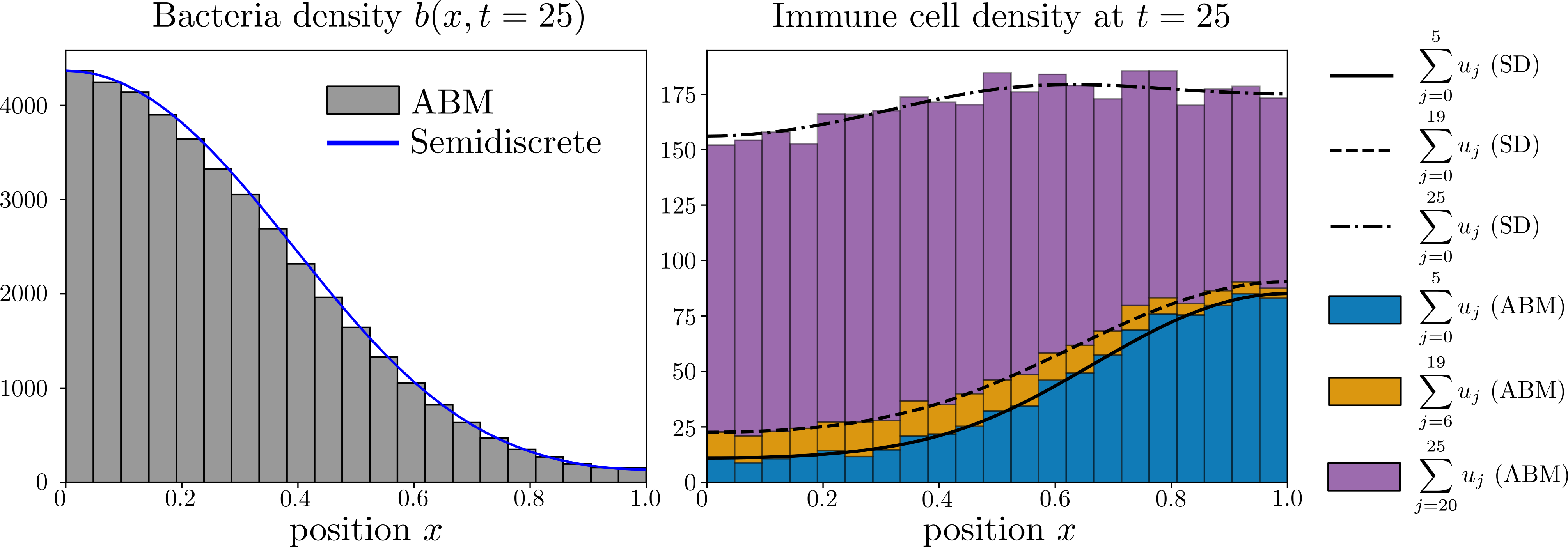}
\caption{Snapshot  of cell and bacteria densities (left and right panels, respectively) at time $t=25$ showing excellent agreement between ABM and the semidiscrete (SD) system. The fraction of the immune cells in various infection states is shown as stacked bars and different line styles.  Results from the ABM were obtained by ensemble averaging  120 independent simulations. }
\label{fig:eg1_ub_snapshot}
\end{figure}

We plot the number of bacteria and immune cells in Figure~\ref{fig:eg1_ub}
as well as a snapshot of the immune cell and bacteria densities in
Figure~\ref{fig:eg1_ub_snapshot}. The total bacterial population and state-structured
immune cell counts from the semidiscrete system closely track the ensemble
averages from the ABM over the full simulation horizon
(Figure~\ref{fig:eg1_ub}). The spatial density profiles at \(t=25\) also agree
well (Figure~\ref{fig:eg1_ub_snapshot}), indicating that the semidiscrete system
captures both population-level and spatial features of the
ABM. This close agreement validates our estimates for the ingestion rates $k_j$ using the Smoluchowski-type calculation in Section \ref{sec:semidiscrete} and validates our approach for the special case where $r_0=0$. 

\subsection{Example 2: Comparison between Semidiscrete and Continuum Systems}

Next, we compare the semidiscrete system in Eq.~\eqref{eqn:semidiscrete_system}  with the continuum system in Eq.~\eqref{eqn:continuum_system}. We begin by analysing the steady-states of both models, and then compare transient dynamics with full numerical simulations. 

We analyse the spatially uniform steady-state of the continuum system in Eq.\\~\eqref{eqn:continuum_system} first.  The total number of extracellular bacteria $B(t)$, defined by Eq.~\eqref{eqn:total_counts}, satisfies 
\begin{equation}
	\dfo{B}{t} = -N\int_\Omega\int_0^1 k(s)u(s,\mbf{x},t)b(\mbf{x},t)\,\mathrm{d}\mbf{x}\,\mathrm{d}s \leq 0, 
	\label{eqn:continuum_b_mass_ode}
\end{equation}
where we have used the no-flux boundary conditions in Eq.~\eqref{eqn:continuum_BC_space}. As $u=0$ is ruled out by the boundary condition on $s=0$ in Eq.~\eqref{eqn:continuum_BC_s0_final}, the only non-negative steady state for $b$ from Eq.~\eqref{eqn:continuum_b_mass_ode} and \eqref{eqn:total_counts} is $b=0$.  The steady-state $u^*(s)$ therefore satisfies 
\begin{subequations}
\begin{equation}
	\frac{1}{2N^2}\dft{}{s}\left(f(s,0)u^*\right) - \frac{1}{N}\dfo{}{s}\left(g(s,0) u^*\right) - \beta(s)u^*=0,
	\label{eqn:continuum_ss}
\end{equation}
with boundary conditions 
\begin{align}
	\frac{1}{2N}\dfo{}{s}\left( f(s,0) u^*\right) - g(s,0) u^* &= - a, \quad s=0, \label{eqn:continuum_ss_BC_s0}\\
	\frac{1}{2N}\dfo{}{s}\left( f(s,0) u^*\right) - g(s,0) u^* &=0, \quad s=1.
	\label{eqn:continuum_ss_BC_s1}
\end{align}
\label{eqn:continuum_ss_problem}
\end{subequations}

Next, we analyse Eq.~\eqref{eqn:continuum_ss} in the regime $N\gg 1$. When $s=\mathcal{O}(1)$,  we expand $u^*$ in a regular asymptotic series as 
\begin{equation}
	u^*(s)\sim \sum_{\ell=0}^\infty v_\ell(s) N^{-\ell}.
	\label{eqn:u_outer_expansion}
\end{equation}
We insert \eqref{eqn:u_outer_expansion} into \eqref{eqn:continuum_ss} and equate coefficients at powers of $N^{-1}$. At leading order, we find
\begin{equation}
	\beta(s) v_0 = 0,
\end{equation}
and thus $v_0=0$. Equating coefficients at higher orders yields $v_\ell=0$, $\ell=1,\ldots$, and we conclude that
\begin{equation}
	u^*(s)\sim \est, \quad s=\mathcal{O}(1),
	\label{eqn:u_outer_solution}
\end{equation}
where $\est$ denotes exponentially small terms.  This solution is consistent with the boundary condition \eqref{eqn:continuum_ss_BC_s1} at $s=1$ up to all algebraic orders of $N^{-1}$, but does not satisfy the boundary condition \eqref{eqn:continuum_ss_BC_s0} at leading order. Thus a boundary layer must be present near $s=0$. The appropriate rescaling in this boundary layer that balances all terms in Eq.~\eqref{eqn:continuum_ss}  is
\begin{equation}
	s = N^{-1}S, \quad u(N^{-1}S) = U(S),
	\label{eqn:inner_rescaling}
\end{equation}
where $S=\mathcal{O}(1)$ in the boundary layer. Next we insert the scalings \eqref{eqn:inner_rescaling} into \eqref{eqn:continuum_ss} and expand $f$, $g$, $\beta$, and $U$ as
\begin{align}
	f(N^{-1}S,0) &= f_0 + \mathcal{O}(N^{-1}), \quad g(N^{-1}S,0) = g_0 + \mathcal{O}(N^{-1}),  \nonumber\\
	 \beta(N^{-1}S) &= \beta_0 + \mathcal{O}(N^{-1}),\quad U(S) = U_0(S) + \mathcal{O}(N^{-1}),
	\label{eqn:inner_expansions}
\end{align}
where $f_0 = f(0,0)$, $g_0 = g(0,0)$, and $\beta_0= \beta(0)$. Next we insert \eqref{eqn:inner_expansions} into Eqs.~\eqref{eqn:continuum_ss}--\eqref{eqn:continuum_ss_BC_s0} and equate coefficients of $N^{-1}$. At leading order, this yields
\begin{subequations}
\begin{align}
	\frac{f_0}{2} \dft{U_0}{S} - g_0 \dfo{U_0}{S} - \beta_0 U_0&=0, \quad S>0, \label{eqn:inner_problem_a} \\
	\frac{f_0}{2}\dfo{U_0}{S} - g_0 U_0 &= - a, \quad S=0. \label{eqn:inner_problem_b}
\end{align}
	\label{eqn:inner_problem}	
\end{subequations}
The general solution of \eqref{eqn:inner_problem_a} is 
\begin{equation}
	U_0(S) = A_+e^{p_+S} + A_-e^{p_-S},
\end{equation}
where $A_\pm$ are constants and 
\begin{equation}
	p_\pm = \frac{1}{f_0}\left( g_0 \pm \sqrt{g^2_0+ 2f_0\beta_0}\right).
\end{equation}
Using the definition of $f$ in Eq.~\eqref{eqn:fg_def} we have that $f_0 = q(0) + r(0)>0$; it follows that $p_+>0$ and $p_-<0$. Formally matching with the exponentially small outer solution \eqref{eqn:u_outer_solution} requires that the solution in the boundary layer decays exponentially away from $S=0$; hence we set $A_+=0$. The boundary condition \eqref{eqn:inner_problem_b} then determines the constant $A_-$ as  
\begin{equation}
	A_- = \frac{a}{g_0 - \frac{p_-f_0}{2}} = \frac{2a}{g_0 + \sqrt{g^2_0 + 2f_0\beta_0}}>0.
\end{equation}
The leading order composite solution, valid for all $s\in[0,1]$, is therefore
\begin{equation}
	u^*(s) \sim \frac{2a}{g_0 + \sqrt{g^2_0 + 2f_0\beta_0}}e^{p_-Ns}. \label{eqn:continuum_ss_leading_order}
\end{equation}
Eq.~\eqref{eqn:continuum_ss_leading_order} represents a well-mixed concentration of immune cells whose states are exponentially localized to $s=0$.

Next, we analyse the steady-state of the semidiscrete system in Eq.~\eqref{eqn:semidiscrete_system}, which we denote by $b^*$ and $\mbf{u}^*$. The total number of extracellular bacteria $B(t)$, defined in \eqref{eqn:total_counts}, now satisfies the equation
\begin{equation}
	\dfo{B}{t} = -\sum_{j=0}^{N-1} k_j \int_{\Omega} u_j(\mbf{x},t)b(\mbf{x},t)\,\mathrm{d}\mbf{x} \leq 0,
	\label{eqn:semidiscrete_b_mass_ode}
\end{equation}
which is obtained by integrating \eqref{eqn:b_semidiscrete} and the no-flux boundary conditions \eqref{eqn:continuum_BC_space}. From \eqref{eqn:semidiscrete_b_mass_ode} there are two possibilities for the steady-state: either $b^*=0$ or  $\mbf{u}^* = (0,\ldots,0,u_N^*)^T$. The latter is ruled out because  $(0,\ldots,0,u_N^*)^T$ is not a solution of Eq.~\eqref{eqn:u_semidiscrete} in general. Therefore $b^*=0$ and the spatially uniform steady state $\mbf{u}^*$ is the unique solution of 
\begin{equation}
	\mathcal{M}\mbf{u}^*= -\mbf{a},
	\label{eqn:semidiscrete_ss_exact}
\end{equation}
where we recall that $\mathcal{M}$ is invertible because $\bar{\mathcal{M}}$ defined in Eq.~\eqref{eqn:M_a_bar_def} is invertible. If $r_0=0$, then the exact solution of Eq.~\eqref{eqn:semidiscrete_ss_exact} is simply $\mbf{u}^* = (a/\beta_0,0,\ldots,0)^T$. We will discuss this special case after analysing the general case $r_0>0$.  Eq.~\eqref{eqn:semidiscrete_ss_exact} is exact; however, we wish to compare it with Eq.~\eqref{eqn:continuum_ss_leading_order} in the regime where $N$ is large. 

To accomplish this, we obtain a closed-form expression for the entries of $\mbf{u}^*$ by effectively inverting $\mathcal{M}$ in the regime where $N\gg 1$. This is the discrete analogue of solving \eqref{eqn:continuum_ss_problem}. Using the definition of $\mathcal{M}$ in Eq.~\eqref{eqn:KM_def},   we have 
\begin{subequations}
\begin{align}
	(r_0 + \beta_0)u^*_0 - q_1 u_1^* &= a, \label{eqn:recursion_j0}\\
	-r_{j-1}u^*_{j-1} + \alpha_j u^*_j - q_{j+1}u^*_{j+1}&=0, \quad j=1,\ldots N-1, \label{eqn:recursion_outer}\\
	-r_{N-1}u^*_{N-1} + (q_N + \beta_N)u_N^* &= 0, \label{eqn:recursion_j1}
\end{align}
\label{eqn:recursion_full}
\end{subequations}
where we define $\alpha_j$ by
\begin{equation}
	\alpha_j = q_j + r_j + \beta_j, \quad j=1,\ldots N-1. \label{eqn:alpha_def}
\end{equation}
   Working analogously to the continuum system, we observe that $u_j^*=0$ satisfies both \eqref{eqn:recursion_outer} and \eqref{eqn:recursion_j1}, but does not satisfy \eqref{eqn:recursion_j0}.  We therefore expect
\begin{equation}
	u_j^* \sim \est, \quad j=\mathcal{O}(N).
	\label{eqn:uj_outer_solution}
\end{equation}
To satisfy \eqref{eqn:recursion_j0}, we consider a discrete boundary layer in the region $0\leq j \ll N$ in which $u_j^*$ varies rapidly. 

To analyze the solution in the discrete boundary layer $0\leq j\ll N$, we expand the coefficients using the definitions of the functions $q$, $r$, and $\beta$ in Eq.~\eqref{eqn:continuum_coeffs}. Assuming that the derivatives of these functions are $\mathcal{O}(1)$, we have 
\begin{subequations}
\begin{align}
	r_{j-1} &= r\left(\frac{j-1}{N}\right)\sim r(0) + \mathcal{O}(N^{-1}), \\
	q_{j+1} &= q\left(\frac{j+1}{N}\right) \sim q(0) + \mathcal{O}(N^{-1}), \\
	\alpha_j &= q\left(\frac{j}{N}\right) + r\left(\frac{j}{N}\right) + \beta\left(\frac{j}{N}\right) = q(0) + r(0) + \beta(0) + \mathcal{O}(N^{-1}).
\end{align}
\label{eqn:discrete_coeff_expansions}%
\end{subequations}
We note that $r(0)=r_0$, but there is no counterpart for $q(0)$  because $q_0$ is not a parameter in Eq.~\eqref{eqn:semidiscrete_system}. To avoid confusion, we use the function notation $q(0)$ rather than subscript `$0$'. Next, we insert the expansions \eqref{eqn:discrete_coeff_expansions} into the recursion relation \eqref{eqn:recursion_outer} to obtain, at leading order,
\begin{equation}
	-r(0) u_{j-1}^*  + \alpha(0)u_j^* - q(0) u_{j+1}^*=0, \quad 1\leq j\ll N, \label{eqn:recursion_inner}
\end{equation}
where $\alpha(0) = q(0) + r(0) + \beta(0)$.  Eq.~\eqref{eqn:recursion_inner} is a constant coefficient recursion relation; the general solution is 
\begin{subequations}
\begin{align}
	u_j^* &= \tilde{A}_- \gamma_-^j + \tilde{A}_+\gamma_+^j, \quad  0\leq j\ll N, \label{eqn:uj_inner_solution_general}\\
	\gamma_\pm &= \frac{1}{2q(0)}\left(\alpha(0) \pm \sqrt{(\alpha(0))^2 - 4q(0)r(0)}\right)>0.\label{eqn:gamma_pm_def}
\end{align}
\end{subequations}
It follows from the definition of $\alpha(0)$ in Eq.~\eqref{eqn:alpha_def} that $\alpha(0)^2-4q(0)r(0)>(r(0)-q(0))^2$; hence $\gamma_+>1$. For the other root, we rearrange Eq.~\eqref{eqn:gamma_pm_def} as
\begin{equation}
	\gamma_- = \frac{2r}{\alpha + \sqrt{(r - q - \beta)^2 + 4\beta r}} < \frac{2r}{\alpha + |r - q - \beta|} = \frac{2r}{r + q + \beta + |r - q - \beta|} \leq 1,
\end{equation}
where we suppress the evaluation at $s=0$ for brevity. Thus the first term in Eq.~\eqref{eqn:uj_inner_solution_general} decays exponentially away from $j=0$ and the second term grows exponentially. To match $u_j^*$ in Eq.~\eqref{eqn:uj_inner_solution_general} with the exponentially small solution for $j=\mathcal{O}(N)$ in \eqref{eqn:uj_outer_solution}, we demand that $u_j^*\sim \est$ for $j=\mathcal{O}(N)$ in \eqref{eqn:uj_inner_solution_general}. This yields $\tilde{A}_+=0$. To find $\tilde{A}_-$, we insert \eqref{eqn:uj_inner_solution_general} into \eqref{eqn:recursion_j0} to find
\begin{equation}
	\tilde{A}_-=\frac{a}{r(0) + \beta(0) - q_1\gamma_-} \sim \frac{a}{r(0) + \beta(0) - q(0)\gamma_-} + \mathcal{O}(N^{-1}).
\end{equation} 
The leading order composite solution, valid for all $j=0,\ldots, N$, is therefore
\begin{equation}
	u_j^*\sim \frac{a}{r(0) + \beta(0) - q(0)\gamma_-}\gamma_-^j ,\quad j=0,\ldots, N.
	\label{eqn:semidiscrete_ss_leading_order_temp}
\end{equation}
For ease of comparison with the continuum system, we rewrite Eq.~\eqref{eqn:semidiscrete_ss_leading_order_temp} using the definition of $\gamma_-$ in  Eq.~\eqref{eqn:gamma_pm_def} as well as the definitions of $f$ and $g$ in Eq.~\eqref{eqn:fg_def}. This yields
\begin{equation}
	u_j^*\sim \frac{2a}{g_0 + \sqrt{g_0^2 + 2f_0\beta_0 + \beta_0^2} + \beta_0} e^{(\ln\gamma_-)j}.
	\label{eqn:semidiscrete_ss_leading_order}
\end{equation}
Eq.~\eqref{eqn:semidiscrete_ss_leading_order} is the counterpart of \eqref{eqn:continuum_ss_leading_order} for the semidiscrete system \eqref{eqn:semidiscrete_system}. 

Comparing the two steady-state solutions in Eq.~\eqref{eqn:continuum_ss_leading_order} and Eq.~\eqref{eqn:semidiscrete_ss_leading_order}, we observe that the decay rates and the coefficients differ in general. Disagreement between discrete and continuum models within boundary layers is a known phenomenon observed previously in similar models (see \cite{agostinelli2026multiscale} and the references therein). Even though the solutions differ in the boundary layer, the total number of cells at steady-state agrees at leading order. To see this explicitly, we sum \eqref{eqn:semidiscrete_ss_leading_order} over $j$ and integrate \eqref{eqn:continuum_ss_leading_order} over $s$ to find, after some algebra
\begin{subequations}
\begin{align}
	\sum_{j=0}^Nu_j^* & \sim \frac{a}{\beta_0}, \label{eqn:sd_cell_total_leading_order}\\
	N\int_0^1 u^*(s)\,\mathrm{d}s &\sim \frac{a}{\beta_0}.
\end{align}
\label{eqn:ss_mass}%
\end{subequations}
For the special case where $r_0=0$, we have $\mbf{u}^*=(a/\beta_0,0,\ldots,0)^T$, so the boundary layer is  effectively confined to  the $j=0$ state only and  Eq.~\eqref{eqn:sd_cell_total_leading_order} is actually exact. However, the conclusion of differing boundary layer structures for the semidiscrete and continuum models still applies in this special case.

Having characterised the steady-states of the semidiscrete system \eqref{eqn:semidiscrete_system} and continuum system \eqref{eqn:continuum_system} analytically, we now compare the full spatio-temporal dynamics \emph{via} numerical solution of both systems.  We select parameters that meet the smoothness assumptions underlying the derivation of Eq.~\eqref{eqn:continuum_system}. For the continuum system, we use 
\begin{subequations}
\begin{align}
	D(s) &= 0.01\left(1 -\frac{s}{2}\right), \quad r(s) = 1-\frac{s}{2}, \quad q(s) = \frac{3}{2} + \frac{s}{2},\nonumber \\ k(s) &= 0.01\left(1 - \frac{s^2}{2}\right), \quad \beta(s) = 0.1 e^{s}, \quad D_b = 10^{-3}, \quad a=1,
	\label{eqn:eg2_params}
\end{align}
and we use the following initial condition
\begin{align}
	u(s,\mbf{x},0) &= \frac{25}{N\sqrt{2\pi \delta^2}}\exp\left(-8\left(s-\frac{1}{2}\right)^2 - \frac{\left(x-\frac{3}{4}\right)^2}{2\delta^2}\right), \\
	b(\mbf{x},0)&= \frac{100}{\sqrt{2\pi \delta^2}}\exp\left(- \frac{\left(x-\frac{1}{4}\right)^2}{2\delta^2}\right), 
	\label{eqn:eg2_IC}
\end{align}
\label{eqn:eg2_vals}%
\end{subequations}
where $\delta^2=0.01$ and we consider one spatial dimension for simplicity, i.e., $\Omega = (0,1)$ and $\mbf{x}=x$. For the semidiscrete system, we simply evaluate each of the functions in Eq.~\eqref{eqn:eg2_vals} at $s_j=j/N$, $j=0,\ldots, N$. We note that the values of $k_j=k(s_j)$ are not computed from \eqref{eqn:effective_reaction_rate} because the ABM is not used in this example.

\begin{figure}
\centering
\includegraphics[width=0.95\textwidth]{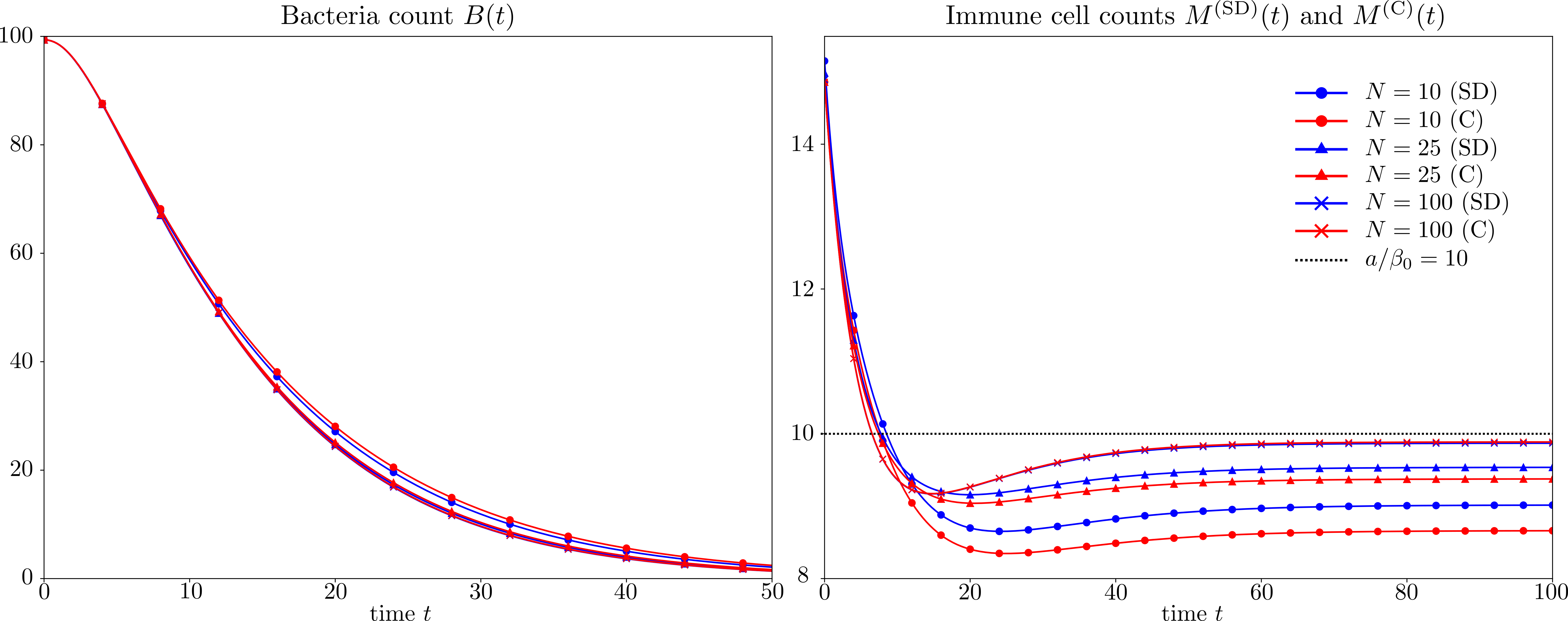}
\caption{Comparison of the semidiscrete (SD) system in Eq.~ \eqref{eqn:semidiscrete_system} and continuum (C) system in Eq.~\eqref{eqn:continuum_system} for different values of $N$. Left: Total bacteria population $B(t)$ defined in  Eq.~\eqref{eqn:total_counts}. Right: Total immune cell population defined by Eq.~\eqref{eqn:total_counts}.   }
\label{fig:eg2_ub}
\end{figure}

\begin{figure}
\centering
\includegraphics[width=0.95\textwidth]{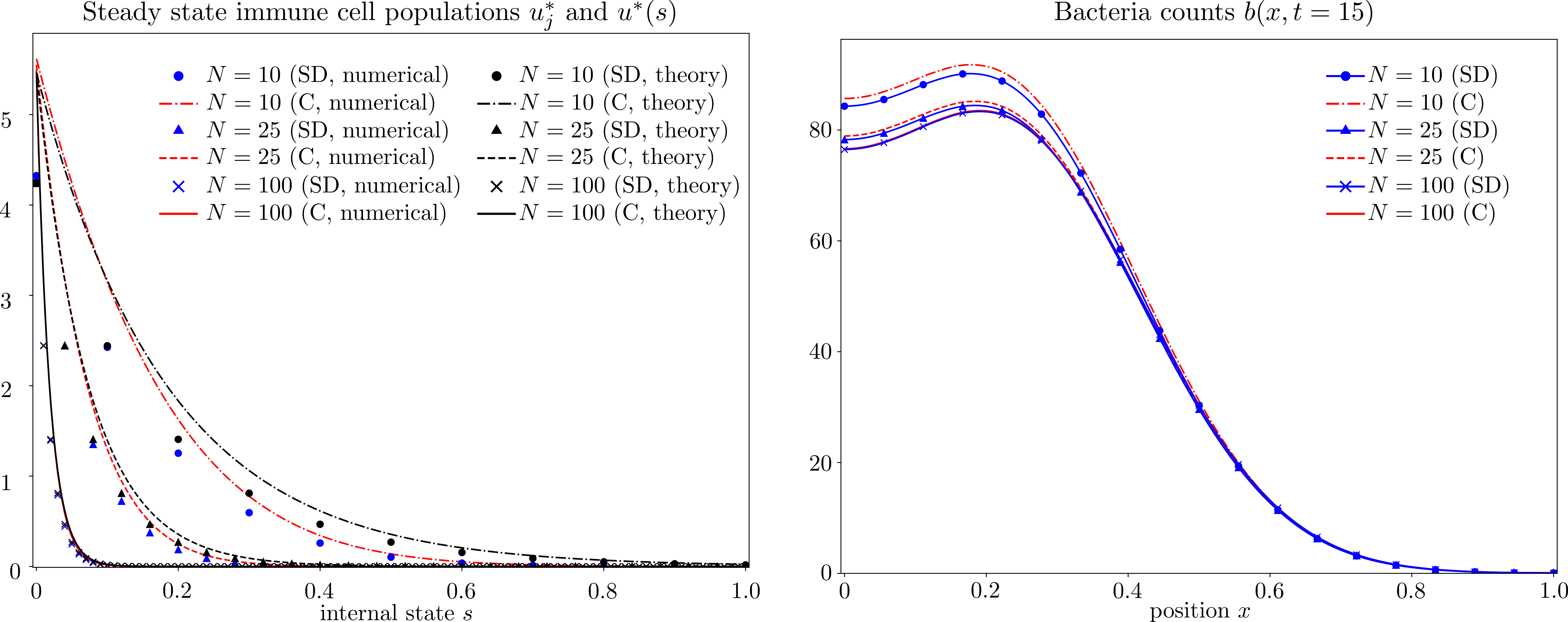}
\caption{Comparison of the semidiscrete (SD) system in Eq.~ \eqref{eqn:semidiscrete_system} and continuum (C) system in Eq.~\eqref{eqn:continuum_system} for different values of $N$. Left: Approximate steady-state immune cell populations in each state $U_j$ and $U(s,t)$ at $t=100$. Our leading order predictions from the boundary layer analysis in Eqs.~\eqref{eqn:continuum_ss_leading_order}, \eqref{eqn:semidiscrete_ss_leading_order} are shown in black. Right: Snapshot of the bacteria population $b(x,t)$ at $t=15$. }
\label{fig:eg2_ub_snapshot}
\end{figure}

We plot the total cell populations $M^{(\mathrm{SD})}(t)$ and
$M^{(\mathrm{C})}(t)$, defined in Eq.~\eqref{eqn:total_counts}, in
Figure~\ref{fig:eg2_ub}. We observe excellent agreement in the total bacteria
population even at a relatively small value of $N=10$. Close agreement is also
present in the spatial distribution of bacteria shown in
Figure~\ref{fig:eg2_ub_snapshot}. At larger values of $N$, the total cell counts
agree reasonably well (Figure~\ref{fig:eg2_ub}). We also observe good agreement
with our theoretical predictions in
Eqs.~\eqref{eqn:continuum_ss_leading_order},
\eqref{eqn:semidiscrete_ss_leading_order}, and \eqref{eqn:ss_mass} from our
steady-state analysis (Figure~\ref{fig:eg2_ub}, right panel, and
Figure~\ref{fig:eg2_ub_snapshot}, left panel). As predicted, the solutions
differ in the boundary layer near $s=0$.

This agreement can be made explicit at steady state. Because both the
semidiscrete and continuum steady states are exponentially localized in a boundary layer near
$s=0$, and because $k$ varies slowly there, $k_j=k(0)+\mathcal O(N^{-1})$
and $k(s)=k(0)+\mathcal O(N^{-1})$ throughout the region $0\leq j \leq N'$, where $1\ll N'\ll N$, that contributes at
leading order.  By decomposing the sum $\sum_{j=0}^{N-1}k_j u_j^*$ into regions $j\leq N'$ and $j>N'$, and using Eq.~\eqref{eqn:ss_mass}, we obtain
\begin{align*}
	\sum_{j=0}^{N-1} k_j u_j^* & \sim \sum_{j=0}^{N'}k_j u_j^* + \est \sim k(0)\sum_{j=0}^{N'} u_j^* + \mathcal{O}(N^{-1})  \\
	&\sim k(0)\sum_{j=0}^{N} u_j^*  + \mathcal{O}(N^{-1}) \sim \frac{k(0)a}{\beta_0}+\mathcal O(N^{-1}).
\end{align*}
By a similar decomposition of the integral $\int_0^1 k(s)u^*(s)\,\mathrm{d}s$ \emph{via} Laplace's method (see e.g., \cite{hinch1991perturbation}, Chapter 3), we obtain
\begin{equation*}
N\int_0^1 k(s)u^*(s)\,\mathrm{d}s
\sim \frac{k(0)a}{\beta_0}+\mathcal O(N^{-1}).
\end{equation*}
Thus the bacterial equations in the semidiscrete and continuum systems agree at
leading order, even though the state distributions differ pointwise inside the
boundary layer.

\subsection{Example 3: Comparison between all Three Models}

In this final example, we compare the ABM, with the semidiscrete and continuum systems in Eqs.~\eqref{eqn:semidiscrete_system} and \eqref{eqn:continuum_system} in a biologically relevant parameter regime, similar to Eq.~\eqref{eqn:eg1_params}. We use the following parameter values
\begin{align}
	q(s) &= 1, \quad r(s) =  5s\left(1 - \frac{Ns}{N+10}\right), \quad \beta(s)= 5\times 10^{-2},   \nonumber \\
	D(s) &= 5\times 10^{-3}\left(1- \frac{s}{2}\right), \quad D_b = 2\times10^{-3}, \quad a = 10, \quad N=100.
	\label{eqn:eg3_params}
\end{align}
The  parameters $q_j$, $r_j$, $\beta_j$, and $D_j$ are obtained by evaluating the corresponding functions in Eq.~\eqref{eqn:eg3_params} at $s=j/N$. 

We  use the same initial conditions as Example 1 for the ABM and semidiscrete system (see Eq.~\eqref{eqn:eg1_IC}).  For the continuum system, we simply interpolate $U_j(0)$ in Eq.~\eqref{eqn:eg1_IC_u} using the natural cubic spline to obtain the initial condition  $u(s,x,0)$. 

The functional form of the ingestion rates $k_j$ in terms of $j$ is not given explicitly by \eqref{eqn:effective_reaction_rate}. To be able to solve the continuum system \eqref{eqn:continuum_system}, we must be able to evaluate both $k(s)$ and its derivative at arbitrary values of $s\in [0,1]$. To do this, we calculate the discrete values of $k_j$ using \eqref{eqn:effective_reaction_rate} and interpolate the result using the natural cubic spline. Because the index $j$ runs from $0$ to $N-1$ for $k_j$, interpolation alone cannot be used to evaluate $k(s)$ for $s\in (1-1/N, 1]$. To obtain values for $k(s)$ in this small region near $s=1$, we use a simple linear extrapolation from the endpoint of the cubic spline.  We plot the effective ingestion rate coefficients in Figure \ref{fig:eg1_reaction_rates}.

To facilitate comparison between all three models, we define the following immune cell counts
\begin{subequations}
\begin{align}
	u_{\mathrm{low}}^{(\mathrm{SD})} &= \sum_{j=0}^{10}u_j, \quad u_{\mathrm{mid}}^{(\mathrm{SD})} = \sum_{j=11}^{20}u_j, \quad 
	u_{\mathrm{high}}^{(\mathrm{SD})} = \sum_{j=21}^{100}u_j, \\
	u_{\mathrm{low}}^{(\mathrm{C})} &=N\int_0^{0.1}u\,\mathrm{d}s, \quad u_{\mathrm{mid}}^{(\mathrm{C})} =N\int_{0.1}^{0.2}u\,\mathrm{d}s, \quad
	u_{\mathrm{high}}^{(\mathrm{C})} =N\int_{0.2}^{1}u\,\mathrm{d}s,
\end{align}
\label{eqn:low_mid_high_def}%
\end{subequations}
which roughly indicate the number of immune cells with low, medium, and high bacterial loads for the semidiscrete (SD) and continuum (C) systems. The corresponding quantities for the ABMs are calculated by simply ensemble averaging the relevant agent counts.

\begin{figure}
\centering
\includegraphics[width=0.95\textwidth]{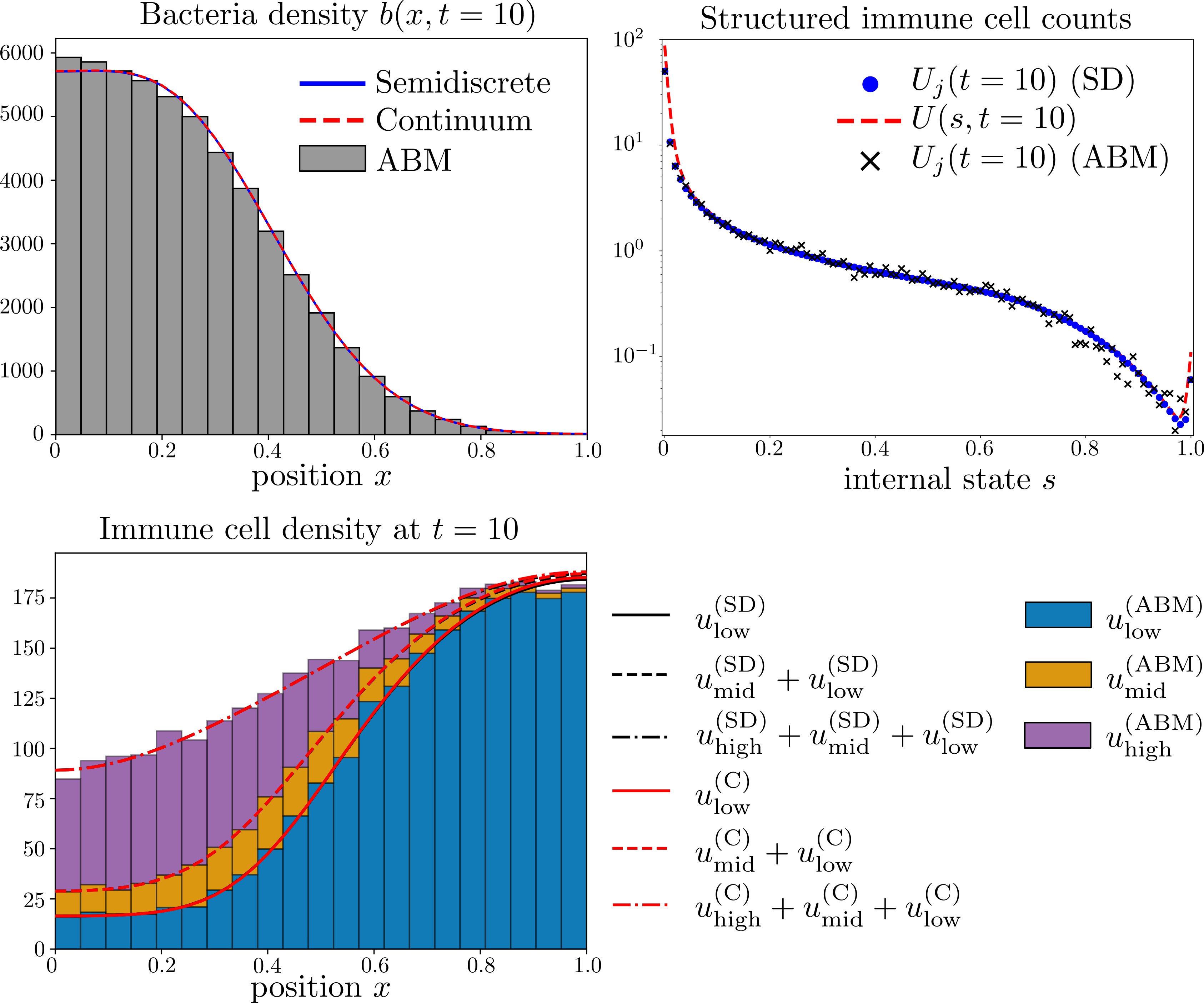}
\caption{Solution snapshots at $t=10$ showing excellent agreement between the
ABM, the semidiscrete (SD) system, and the continuum (C) system. Top left:
bacteria density. Top right: immune-cell counts structured by internal state,
calculated from Eq.~\eqref{eqn:u_counts_def}. Bottom: immune-cell densities
grouped by bacterial load according to Eq.~\eqref{eqn:low_mid_high_def}.}
\label{fig:eg3_ub_snapshot}
\end{figure}

\begin{figure}
\centering
\includegraphics[width=0.95\textwidth]{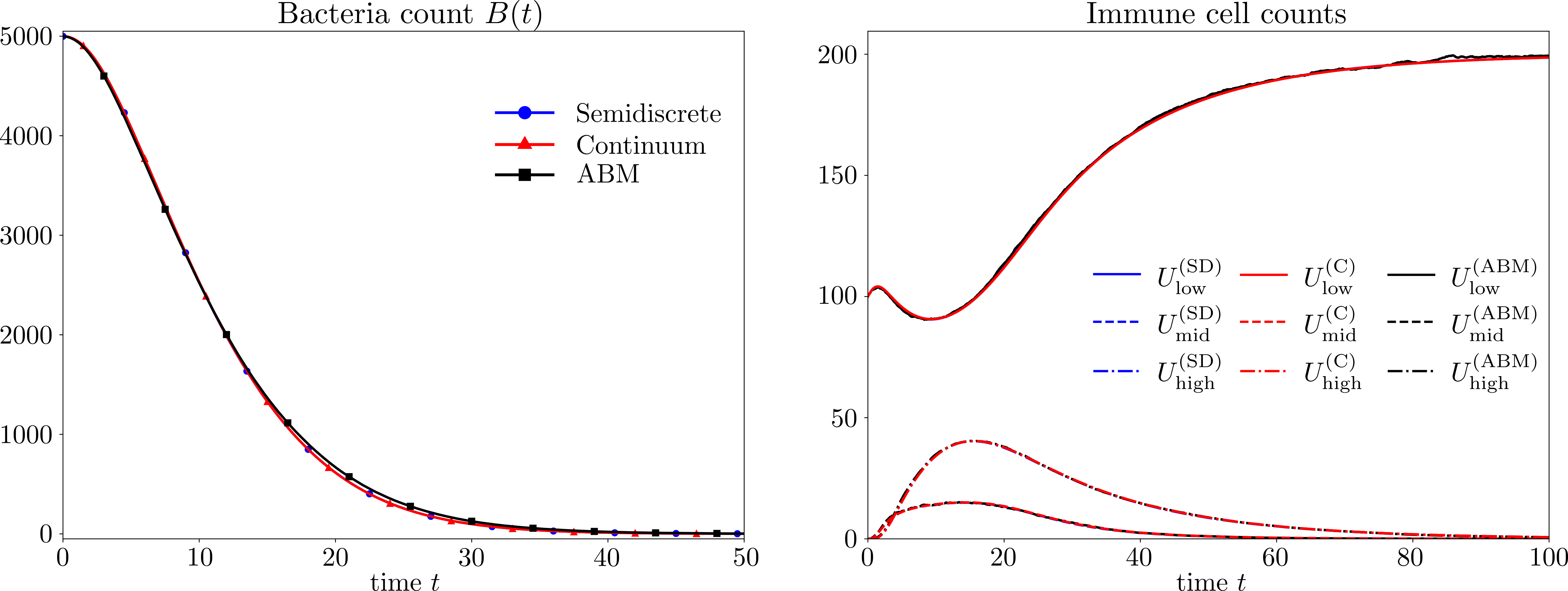}
\caption{Comparison of total number of bacteria (left) and immune cells (right) for the ABM, semidiscrete system, and the continuum system showing excellent agreement. Cell counts $U_{\mathrm{low}}$, $U_{\mathrm{mid}}$, $U_{\mathrm{high}}$ are calculated by integrating Eq.~\eqref{eqn:low_mid_high_def} over space. }
\label{fig:eg3_ub}
\end{figure}

 We compare the predictions of all three models in Figures \ref{fig:eg3_ub_snapshot} and \ref{fig:eg3_ub}. Figure \ref{fig:eg3_ub_snapshot} shows that both the semidiscrete and continuum systems reproduce the ensemble-averaged spatio-temporal  features of the ABM as well as the distribution of internal states. In both figures, the semidiscrete and continuum solutions are almost visually  indistinguishable. 

\section{Discussion} \label{sec:discussion}

In this work, we constructed an ABM of a within-host bacterial infection motivated by frequently occurring rules in the literature. We showed how these rules can be coarse-grained into an effective \emph{semidiscrete} reaction-diffusion system, which is structured by a discrete state variable representing the number of ingested bacteria. This coarse graining involved a Smoluchowski-type analysis of an auxiliary diffusion problem to compute effective ingestion rates $k_j$. We  observed excellent agreement between the ABM and the semidiscrete system using our computed ingestion rates. In the limit of large intracellular capacity $N\gg 1$, we showed how the semidiscrete system can be further coarse grained into a \emph{continuum} system whose states are structured by a continuous variable. We analysed the steady states of the continuum and semidiscrete systems and found that the solutions differ within boundary layers, although full numerical solutions still show good qualitative and quantitative agreement. 

In calculating the effective ingestion rates $k_j$, we analysed the steady
state of an auxiliary diffusion problem. A full characterization of the
assumptions underlying this Smoluchowski-type calculation is beyond the scope of
this work, but the main requirements are that the system is dilute, that
interactions between distinct ingestion events are negligible, and that the
local diffusion problem relaxes on a time scale short compared with macroscopic
changes in the cell and bacterial concentrations. One might therefore expect the
approximation to be most accurate when the intracellular state distribution is
close to local quasi-equilibrium. The numerical examples are more encouraging:
even during transient regimes in which the state distributions are not at
equilibrium (Figures~\ref{fig:eg1_ub_snapshot} and
\ref{fig:eg3_ub_snapshot}), the derived $k_j$ lead to close agreement with the
ensemble-averaged ABM results. This suggests that the effective ingestion rates
are robust over a wider range of conditions than the formal derivation alone
would guarantee.

Several extensions and modifications of our model are possible.  Chemical signalling plays a critical role in affecting agent behaviours in  many ABMs \cite{segovia2004identifying, petrucciani2024silico, hoerter2025timing, cilfone2013multi, bowness2018modelling, ray2009synergy, dancik2010parameter, gopalakrishnan2013using, shi2016agent}; therefore, an important extension of our work is to include chemical signalling in our coarse graining. Many ABMs use a
generalized version of our ingestion rule where bacteria agents are ingested
with a finite probability per unit time $\lambda$ while within a distance
$\bar{\rho}$ of an immune cell agent. Here, we effectively assume a limiting
case where $\lambda\to \infty$, but our calculation of the ingestion rates
generalizes straightforwardly (see \cite{erban2020stochastic}). In our ABM, new immune cells are introduced into the system with a position that is drawn from a uniform distribution over the domain. This effectively treats the entire domain as a single compartment with a source rate of $a$ cells per unit time, per unit volume. This could be generalized by dividing the domain into regions with different source rates $a(\mbf{x})$. In principle, our coarse graining of the effective ingestion rates would remain unchanged as long as $a(\mbf{x})$ varies slowly with $\mbf{x}$. The ingestion rates $k_j$ would not depend on $\mbf{x}$ because  Eq.~\eqref{eqn:effective_reaction_rate} turns out to be independent of $a$. 

In our analysis of the semidiscrete and continuum systems, we discovered that boundary layers can form, causing localized disagreement between the two models. This phenomenon has been observed previously in other discrete state-structured models \cite{agostinelli2026multiscale}. Although this disagreement did not appear significant in our examples, it may be problematic in more complex systems. Additionally, the effective ingestion rates $k_j$ seem to have boundary layers near $s=1$ (see Figure \ref{fig:eg1_reaction_rates}) in some parameter regimes. It may be interesting to investigate the boundary layer structure analytically by considering the large $N$ limit of Eq.~\eqref{eqn:effective_reaction_rate}. We may be able to resolve the discrepancy between the two models by utilizing a hybrid approach where the semidiscrete model is used within the boundary layers and coupled to the continuum system in the outer region. Another possibility is to modify our coarse graining to incorporate effective boundary conditions \emph{via} a modified multiple-scales approach \cite{agostinelli2026multiscale}. 

ABMs are often computationally expensive for large systems with many agents, but
they retain the cell-level rules and stochastic heterogeneity needed for detailed
biological modelling. PDE models are computationally cheaper, but describe the
ensemble-averaged behaviour of the system. The coarse-graining approach developed
here suggests a practical way to combine these strengths. For example, the PDE
model could be used for broad parameter exploration, sensitivity analysis, or
initial stages of parameter estimation, after which selected parameter regimes
could be refined with the original ABM. The close agreement observed in our
numerical examples indicates that this multistage workflow is a promising
direction for future work.

\appendix
\section{An Example where the Leading Order Continuum System is Insufficient}
\label{app:leading_order_insufficient}

In this appendix we show a special case where neglecting the $\mathcal{O}(N^{-1})$ boundary terms in \eqref{eqn:continuum_system_b} results in significant disagreement with the semidiscrete system \eqref{eqn:semidiscrete_system} as well as the ABM in Eqs.~\eqref{eqn:ABM_rule_ingestion}--\eqref{eqn:ABM_rules_position}.  This special case occurs when all of the rules in the ABM are switched off, except for ingestion of bacteria, i.e., $q_j=r_j=\beta_j=a=0$. Intuitively, the maximum possible number of extracellular bacteria that can be
ingested is simply $M\cdot N$, where $M$ is the number of immune cells. This is because immune cells are neither added nor removed from the system (because $a=\beta_j=0$)  and the intracellular state can only change \emph{via} ingestion. Hence there are two possible outcomes. Either there are too many bacteria for the immune cells to ingest, resulting in all immune cells eventually saturating in the $j=N$ state with remaining extracellular bacteria; or there are sufficiently many immune cells to ingest all extracellular bacteria resulting in a steady, non-saturated distribution of immune cell states. We will see that the second possibility does not occur if the higher order terms are neglected in the continuum system.

We start by considering the steady-state semidiscrete system \eqref{eqn:semidiscrete_system} with $q_j=r_j=\beta_j=a=0$. Denoting the steady-state bacteria and immune cell concentrations as $b^*$ and $u_j^*$, we have from Eq.~\eqref{eqn:semidiscrete_system}
\begin{subequations}
\begin{align}
	\left(\mathcal{K}\mbf{u}^*\right)b^*&=\mbf{0}, \\
	b^*\sum_{j=0}^{N-1} k_j u_j^* &=0,
\end{align}
\end{subequations}
 where we consider a well-mixed system for simplicity. Restricting our attention to the non-trivial, non-negative steady-states, there are two possibilities. Either $b^*=0$ and $\mbf{u}^*$ is arbitrary, or $b^*>0$ and $\mbf{u}^*\in \mathrm{null}(\mathcal{K}) = \mathrm{span}\{(0,\ldots, 1)^T\}$. If there are $M$ immune cells in the system, then $\mbf{u}^*=(0,\ldots,0, M)^T$. These two cases correspond to the two situations described above for the ABM. The former  corresponds to the situation where all the bacteria have been ingested and the distribution of immune cell states is fixed in time. The latter case corresponds to the situation where there are too many bacteria to ingest and the immune cells are completely saturated.

Next we consider the continuum system \eqref{eqn:continuum_system} at steady-state with $q(s)=r(s)=\beta(s)=a=0$. However, we retain the $\mathcal{O}(N^{-1})$ terms in the approximation of the sum in Eq.~\eqref{eqn:integral_approx}, which when inserted into Eq.~\eqref{eqn:b_eqn_rewrite} leads to the system 
\begin{subequations}
\begin{align}
	0 &=  \frac{b^*}{2N^2}\dft{}{s}\left(k(s)u^*\right) - \frac{b^*}{N}\dfo{}{s}\left(k(s)u^*\right),\quad  (s,\mbf{x})\in (0,1)\times \Omega,\label{eqn:continuum_system_simple_u}\\
	0 &=  b^* \left(N\int_0^1 k(s)u^*(s,\mbf{x})\,\mathrm{d} s - \frac{1}{2}k(s) u^*(s,\mbf{x})\bigg|_{s=0}^{s=1}\right), \quad \mbf{x} \in \Omega, \label{eqn:continuum_system_simple_b}
\end{align}
with no-flux boundary conditions given by
\begin{equation}
	\left(\frac{1}{2N}\pfo{}{s}\left( k(s) u^*\right) - k(s) u^*\right)b^* =0, \quad s=0,1,\label{eqn:continuum_system_simple_u_BC_s}
\end{equation}
\label{eqn:continuum_system_simple}%
\end{subequations}
where we again assume a well-mixed system for simplicity. The  reason that the additional boundary terms are needed in  Eq.~\eqref{eqn:continuum_system_simple_b} for this special case is because the RHS of Eq.~\eqref{eqn:continuum_system_simple_b} is  the integral of the state flux in Eq.~\eqref{eqn:continuum_system_simple_u_BC_s} and will therefore turn out to be automatically zero   at steady-state, even when  $b^*$ is non-zero. To show this, we first observe that $b^*$ multiplies every term in Eq.~\eqref{eqn:continuum_system_simple}. Hence there is a non-trivial steady state given by $b^*=0$ and $u^*= u^*(s)$ arbitrary. This corresponds to the situation described above where all bacteria have been ingested and there is some distribution of non-saturated immune cell states. This steady-state would be present even if we neglected the boundary terms in \eqref{eqn:continuum_system_simple_b}. 

\begin{figure}
\centering 
\includegraphics[width=0.5\textwidth]{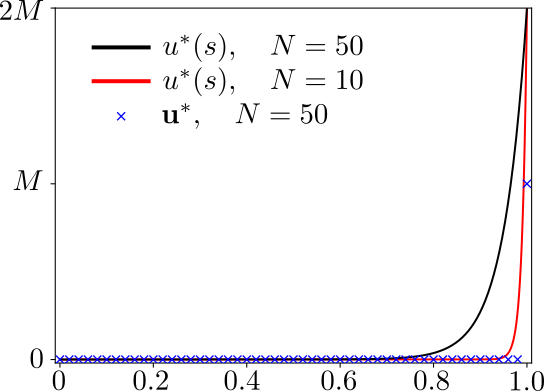}
\caption{Steady-state solutions of the semidiscrete system $\mbf{u}^*=(0,\ldots,0,M)^T$ and continuum system $u^*(s)$ given in \eqref{eqn:continuum_system_simple_ss} for $N=10$ and $N=50$. We use $k(s)=1$. }
\label{fig:appendix_ss}
\end{figure}

If $b^*>0$, then Eq.~\eqref{eqn:continuum_system_simple_u} can be solved directly to find 
\begin{equation}
	u^*(s) = \frac{A}{k(s)} e^{2Ns},
	\label{eqn:continuum_system_simple_ss_temp}
\end{equation}
where the normalization  constant $A$ is found by demanding that the total number of immune cells is  equal to $M$, in particular
\begin{equation}
	M = N\int_0^1 u^*(s)\,\mathrm{d}s  = Ae^{2N}\left(\frac{1}{2 k(1)} + \mathcal{O}(N^{-1})\right),
\end{equation}
where we have obtained the leading order contribution to the integral when $N\gg 1$ using Laplace's method (see e.g.,\cite{hinch1991perturbation}, Chapter 3). Thus to leading order, \eqref{eqn:continuum_system_simple_ss_temp} becomes 
\begin{equation}
	u^*(s) \sim  2 M\frac{ k(1)}{k(s)} e^{2N(s-1)}.
	\label{eqn:continuum_system_simple_ss}
\end{equation}
Substituting the exact expression \eqref{eqn:continuum_system_simple_ss_temp} into Eq.~\eqref{eqn:continuum_system_simple_b} shows that the integral term and endpoint correction cancel exactly, so $b^*>0$ is arbitrary. The steady-state in Eq.~\eqref{eqn:continuum_system_simple_ss} is exponentially small everywhere except within a thin boundary layer of width  $\mathcal{O}(N^{-1})$ near $s=1$. Physically, this steady state represents a situation where there are remaining extracellular bacteria and the immune cell states are  effectively saturated near $j=N$. This is analogous to the steady state of the semidiscrete system where $\mbf{u}^*=(0,\ldots,0,M)^T$, except that the states are concentrated in the boundary layer instead of lying exactly on the $j=N$ boundary. The distribution of internal states \eqref{eqn:continuum_system_simple_ss} can be thought of as a smooth approximation of the steady-state $\mbf{u}^*= ( 0,\ldots,0, M)$ of the semidiscrete system when the number of internal states $N$ is large. We plot the steady-states given by $\mbf{u}^*=(0,\ldots,0,M)^T$ and  Eq.~\eqref{eqn:continuum_system_simple_ss} in Figure \ref{fig:appendix_ss}.

If we had neglected the boundary terms in Eq.~\eqref{eqn:continuum_system_simple_b}, then the only steady-state of the continuum system would be given by $b^*=0$ and $u^*(s)$ arbitrary. This is because we would have found that  Eq.~\eqref{eqn:continuum_system_simple_ss} does not automatically satisfy Eq.~\eqref{eqn:continuum_system_simple_b}. The absence of the steady-state where $b^*>0$ implies that  extracellular bacteria can effectively be removed from the system without being ingested by immune cells, which is inconsistent with both the semidiscrete system and the underlying ABM. We have demonstrated that the boundary term is needed in the special case of a well-mixed, steady-state system where $q=r=\beta=a=0$. In the general case where these parameters are positive, there is no non-zero steady-state for $b^*$, so retaining the higher order terms is not needed for qualitative agreement at steady-state. Moreover, the presence of boundary layers near $s=0$ (see the main text) implies that retaining the additional boundary terms would not result in a higher order approximation of the semidiscrete system. Thus we simply use the leading order term throughout the main text. 

\section{The Ingestion Rates $k_j$ are Regular in the Limit $r_0\to 0^+$} \label{app:kj_regular}

Our goal in this appendix is to show that $k_j=\mathcal{O}(1)$ as $r_0=\varepsilon\to0^+$. Because $\varepsilon>0$, the formula for the ingestion rates in  Eq.~\eqref{eqn:effective_reaction_rate} remains valid. Hence we only need to determine the behaviour as $\varepsilon\to0^+$  of the particular solution $\bar{\mathcal{M}}^{-1}\bar{\mbf{a}}$, the eigenpairs $(\lambda_m,\mbf{e}_m)$ of $\bar{\mathcal{M}}$ in Eq.~\eqref{eqn:M_a_bar_def}, and the coefficients $C_m$. We write $\bar{\mathcal{M}}(\varepsilon)$  to denote the explicit dependence on $\varepsilon$. We note that $\bar{\mathcal{M}}(\varepsilon) = \bar{\mathcal{M}}(0) + \varepsilon \bar{\mathcal{M}}_1$, where 
\begin{equation}
	\bar{\mathcal{M}}_1 = \begin{pmatrix}
		\bar{D}_0^{-1}		& 	0			&	\ldots & 	0 \\
		-\bar{D}_1^{-1}		& 	0			&	\ldots 	& 	0\\
		\vdots	&				&			&	\vdots	\\
		0		&				&			&	0
	\end{pmatrix}.
	\label{eqn:kj_reg_M1_def}
\end{equation}
In the main text, we showed that $\det\bar{\mathcal{M}}(\varepsilon)>0$  for all $\varepsilon>0$. By applying a similar calculation directly to  $\bar{\mathcal{M}}(0)$ (skipping the similarity transform in Eq.~\eqref{eqn:A_similar}), it can be shown that $\det\bar{\mathcal{M}}(0)>0$. The eigenvalues of $\bar{\mathcal{M}}(\varepsilon)$ are therefore $\mathcal{O}(1)$ as $\varepsilon\to 0^+$. 

We begin by considering the eigenpairs of $\bar{\mathcal{M}}(\varepsilon)$, which we denote by $(\lambda_j, \mbf{e}_j)$, $j=0,\ldots, N$.  One of the eigenvalues of $\bar{\mathcal{M}}(0)$ is $\lambda_0^{(0)}=\beta_0\bar{D}_0^{-1}$ and the corresponding eigenvector is $\mbf{e}_0^{(0)}=(1,0,\ldots,0)^T$.  Therefore, when $\varepsilon>0$, the eigenpair $(\lambda_0, \mbf{e}_0)$ is perturbed as
\begin{equation}
		\lambda_0 = \lambda_0^{(0)} + \varepsilon\lambda_0^{(1)} + \mathcal{O}(\varepsilon^2), \quad  \mbf{e}_0 = \mbf{e}_0^{(0)} + \varepsilon\mbf{e}_0^{(1)} + \mathcal{O}(\varepsilon^2).  \label{eqn:kj_reg_eigenpair_expansion}
\end{equation}
Inserting these expansions into the eigenvalue problem $\bar{\mathcal{M}}(\varepsilon)\mbf{e}_0 =\lambda_0\mbf{e}_0$ and equating terms at powers of $\varepsilon$ yields, at order $\mathcal{O}(1)$ and $\mathcal{O}(\varepsilon)$
\begin{subequations}
\begin{align}
	\left(\bar{\mathcal{M}}(0) - \lambda_0^{(0)}I \right)\mbf{e}_0^{(0)}&=\mbf{0}, \label{eqn:kj_reg_eigenpair_O1}\\
	\left(\bar{\mathcal{M}}(0) - \lambda_0^{(0)}I \right) \mbf{e}_0^{(1)} &= \lambda_0^{(1)}\mbf{e}_0^{(0)} - \bar{\mathcal{M}}_1\mbf{e}_0^{(0)}. \label{eqn:kj_reg_eigenpair_Oeps}
\end{align}
\end{subequations}
Eq.~\eqref{eqn:kj_reg_eigenpair_O1} is simply the unperturbed eigenvalue problem for the eigenpair $(\lambda_0^{(0)},\mbf{e}^{(0)}_0)$. By the Fredholm Alternative, Eq.~\eqref{eqn:kj_reg_eigenpair_Oeps} has a solution if, and only if, the RHS is orthogonal to the (one-dimensional) nullspace of $(\bar{\mathcal{M}}(0) - \lambda_0^{(0)}I)^T$. As $\mbf{e}_0^{(0)}$ is also a left eigenvector of $\bar{\mathcal{M}}(0)$, this condition reads
\begin{equation}
	0=\left(\lambda_0^{(1)}\mbf{e}_0^{(0)} - \bar{\mathcal{M}}_1\mbf{e}_0^{(0)}\right)\cdot \mbf{e}_0^{(0)} = \lambda_0^{(1)}-\bar{D}_0^{-1}.
\end{equation}
We demand that a solution of Eq.~\eqref{eqn:kj_reg_eigenpair_Oeps} exists, which implies $\lambda_0^{(1)}=\bar{D}_0^{-1}$. Any solution of  Eq.~\eqref{eqn:kj_reg_eigenpair_Oeps} can be written in the form
\begin{equation}
	\mbf{e}_0^{(1)} = \alpha \mbf{e}_0^{(0)} + \mbf{w}, 
\end{equation}
where $\alpha$ is an arbitrary constant and the entries of $\mbf{w}=(w_0,w_1, w_2,\ldots,w_N)^T$ are given by
\begin{subequations}
\begin{align}
	w_0&=w_1=0, \quad w_2 = -\frac{1}{q_2},\\  w_{j+1}&=\frac{1}{q_{j+1}}\left((r_j + q_j+\beta_j - \beta_0 )w_j - r_{j-1}w_{j-1}\right),\quad  j=2,\ldots, N-1.
\end{align}
\end{subequations}
Thus the eigenvector $\mbf{e}_0$ has the following form
\begin{equation}
	\mbf{e}_0 = (\mathcal{O}(1), \mathcal{O}(\varepsilon^2),\mathcal{O}(\varepsilon), \ldots, \mathcal{O}(\varepsilon))^T, \label{eqn:kj_reg_evec_scaling1}
\end{equation} 
up to an overall normalization.

For the remaining eigenvectors of $\bar{\mathcal{M}}(\varepsilon)$, we will show that each entry is $\mathcal{O}(1)$ as $\varepsilon\to 0^+$ (up to normalization). We do this by  showing that  the  eigenvectors of the unperturbed matrix $\bar{\mathcal{M}}(0)$ have no zero entries, except the eigenvector $\mbf{e}_0^{(0)}=(1,0,\ldots,0)^T$. For a similar result and analysis, see \cite{fernando1997computing} and 4.3.P17 of \cite{horn2012matrix}. We begin by showing that no eigenvector of $\bar{\mathcal{M}}(0)$ can have two consecutive zero entries (except  $\mbf{e}_0^{(0)}$). To see this, we write the components of the unperturbed eigenvalue problem $\bar{\mathcal{M}}(0)\mbf{e}_m^{(0)}=\lambda_m^{(0)}\mbf{e}_m^{(0)}$ explicitly as
\begin{subequations}
\begin{align}
	\left(\frac{\beta_0}{\bar{D}_0} - \lambda_m^{(0)}\right) v_0 - \frac{q_1}{\bar{D}_0} v_1 &= 0, \\
	\left(\frac{r_1 + q_1 +\beta_1}{\bar{D}_1} - \lambda_m^{(0)}\right) v_1  - \frac{q_2}{\bar{D}_1} v_2 &=0, \\
	-\frac{r_{j-1}}{\bar{D}_j}v_{j-1} + \left(\frac{r_j + q_j + \beta_j}{\bar{D}_j} - \lambda_m^{(0)}\right) v_j - \frac{q_{j+1}}{\bar{D}_j} v_{j+1} &=0, \quad j=2,\ldots, N-1, \label{eqn:kj_reg_eigenvector_recurrance_unperturbed_m}\\
	-\frac{r_{N-1}}{\bar{D}_N} v_{N-1} + \left(\frac{q_N + \beta_N}{\bar{D}_N} - \lambda_m^{(0)}\right) v_N &=0, \label{eqn:kj_reg_eigenvector_recurrance_unperturbed_N}
\end{align}
\label{eqn:kj_reg_eigenvector_recurrance_unperturbed}%
\end{subequations}
where we denote the $j^{\text{th}}$ component of $\mbf{e}_m^{(0)}$ as $v_j$ for brevity. If $v_{j'}=v_{j'+1}=0$, where $j'\neq 1$, then Eqs.~\eqref{eqn:kj_reg_eigenvector_recurrance_unperturbed} imply that $v_j=0$ for all $j=1,\ldots, N$. Hence, the only eigenvector with consecutive zero entries is $(1,0,0,\ldots)^T$. Next, suppose that only $v_{j'}=0$ for some $j'$. The cases  $j'=0,N$ lead to $v_j=0$ for all $j$ by Eq.~\eqref{eqn:kj_reg_eigenvector_recurrance_unperturbed} and are therefore excluded. The case $j'=1$ is also excluded as this just reproduces the known eigenpair $(\lambda_0^{(0)}, \mbf{e}_0^{(0)})$. For the general case $j'\neq0,1,N$, we write the eigenvector as $\mbf{e}_m^{(0)}=(\mbf{v}_{\mathrm{t}},0,\mbf{v}_{\mathrm{b}})^T$ and we find 
\begin{subequations}
\begin{align}
	\bar{\mathcal{M}}_{\mathrm{t}} \mbf{v}_{\mathrm{t}} &= \lambda_m^{(0)}\mbf{v}_t, \label{eqn:kj_reg_simultaneous_eigenvals_a}\\
	\bar{\mathcal{M}}_{\mathrm{b}}\mbf{v}_{\mathrm{b}} &= \lambda_m^{(0)}\mbf{v}_{\mathrm{b}}, \label{eqn:kj_reg_simultaneous_eigenvals_b}\\
	r_{j'-1}v_{j'-1} &+ q_{j'+1}v_{j'+1}=0, \label{eqn:kj_reg_simultaneous_eigenvals_fix}
\end{align}
\label{eqn:kj_reg_simultaneous_eigenvals}%
\end{subequations}
where $\bar{\mathcal{M}}_{\mathrm{b}}$ is the matrix obtained by deleting the rows and columns of $\bar{\mathcal{M}}(0)$ up to and including $j'$ (indexed from 0); and  $\bar{\mathcal{M}}_{\mathrm{t}}$ is the matrix obtained by deleting the rows and columns of $\bar{\mathcal{M}}(0)$ including and beyond $j'$. Eq.~\eqref{eqn:kj_reg_simultaneous_eigenvals} states that $\lambda_m^{(0)}$ must  be an eigenvalue of $\bar{\mathcal{M}}_{\mathrm{t}}$ and $\bar{\mathcal{M}}_{\mathrm{b}}$ simultaneously; the condition \eqref{eqn:kj_reg_simultaneous_eigenvals_fix} simply fixes the relative normalizations of $\mbf{v}_{\mathrm{t}}$ and $\mbf{v}_{\mathrm{b}}$.

In general, the eigenvalue $\lambda_m^{(0)}$ will not satisfy  both \eqref{eqn:kj_reg_simultaneous_eigenvals_a} and \eqref{eqn:kj_reg_simultaneous_eigenvals_b}; hence all entries of $\mbf{e}_m^{(0)}$ are non-zero in general. This implies that the remaining eigenvectors of $\bar{\mathcal{M}}(\varepsilon)$ have the scaling
\begin{equation}
	\mbf{e}_m = \left(\mathcal{O}(1), \ldots, \mathcal{O}(1)\right)^T, \quad m=1,\ldots, N. \label{eqn:kj_reg_evec_scaling2}
\end{equation}

Next we consider the particular solution. Denoting  $\mbf{y}=\bar{\mathcal{M}}^{-1}\bar{\mbf{a}}$, we have
\begin{equation}
	\bar{\mathcal{M}}(\varepsilon)\mbf{y} = \bar{\mbf{a}}. 
	\label{eqn:kj_reg_particular_problem}
\end{equation}
The matrix $\bar{\mathcal{M}}$ is invertible when $\varepsilon=0$, so we expand the unique solution $\mbf{y}$ as
\begin{equation}
	\mbf{y} = \mbf{y}_0 + \varepsilon \mbf{y}_1 + \mathcal{O}(\varepsilon^2).
	\label{eqn:kj_reg_y_expand}
\end{equation}
Then inserting Eq.~\eqref{eqn:kj_reg_y_expand} into Eq.~\eqref{eqn:kj_reg_particular_problem} and equating terms at powers of $\varepsilon$, we obtain at $\mathcal{O}(1)$ and $\mathcal{O}(\varepsilon)$
\begin{subequations}
\begin{align}
	\bar{\mathcal{M}}(0) \mbf{y}_0 &= \bar{\mbf{a}}, \label{eqn:kj_reg_particular_problem_O1}\\
	\bar{\mathcal{M}}(0) \mbf{y}_1 &= -\bar{\mathcal{M}}_1\mbf{y}_0.  \label{eqn:kj_reg_particular_problem_Oeps}
\end{align}
\end{subequations}
As $\bar{\mbf{a}}=-(a/\bar{D}_0,0,\ldots,0)^T$ is an eigenvector of $\bar{\mathcal{M}}(0)$ with eigenvalue $\beta_0/\bar{D}_0$, Eq.\\\eqref{eqn:kj_reg_particular_problem_O1} yields
\begin{equation}
	\mbf{y}_0 = -\left(a/\beta_0, 0, \ldots,0\right)^T.
	\label{eqn:kj_reg_y0_soln}
\end{equation}
Then using  Eqs.~\eqref{eqn:kj_reg_M1_def} and  \eqref{eqn:kj_reg_y0_soln} in Eq.~\eqref{eqn:kj_reg_particular_problem_Oeps} gives
\begin{equation}
	\mbf{y}_1 = -\frac{a}{\beta_0}\left(\bar{\mathcal{M}}(0)\right)^{-1}\left(\bar{D}_0^{-1},-\bar{D}_1^{-1},0\ldots,0\right)^T.
\end{equation}
Because the inverse of a general tridiagonal matrix is dense, the vector $\mbf{y}_1$ will generally  have only non-zero entries. Thus the particular solution $\bar{\mathcal{M}}^{-1}\bar{\mbf{a}}$ has the following form 
\begin{equation}
	\bar{\mathcal{M}}^{-1}\bar{\mbf{a}} = \left(\mathcal{O}(1), \mathcal{O}(\varepsilon), \ldots, \mathcal{O}(\varepsilon)\right)^T. \label{eqn:kj_reg_particular_scaling}
\end{equation}

Next, we determine the scaling of the constants $C_0,\ldots,C_N$ and $C$ from the linear system in Eqs.~\eqref{eqn:b_frame_linear_system1} and \eqref{eqn:b_frame_linear_system2}. Using the scalings for the eigenvectors in Eqs.~\eqref{eqn:kj_reg_evec_scaling1}, \eqref{eqn:kj_reg_evec_scaling2}, and particular solution \eqref{eqn:kj_reg_particular_scaling}, we find that the linear system has the following structure
\begin{equation}
	\begin{pmatrix}
		\mathcal{O}(1)				&	\mathcal{O}(1) 	&	 \ldots		&	\mathcal{O}(1)	&	0\\
		\mathcal{O}(\varepsilon^2)	&	 \mathcal{O}(1)	&	\ldots 		&	\mathcal{O}(1)	& 	0 \\
		\vdots						&					&				&					&	\vdots \\
		\mathcal{O}(\varepsilon)		&	\mathcal{O}(1)	&	\ldots		&	\mathcal{O}(1)	&	\mathcal{O}(1)	\\
		\mathcal{O}(\varepsilon)		&	\mathcal{O}(1)	&	\ldots		&	\mathcal{O}(1)	&	\mathcal{O}(1)		
	\end{pmatrix}
	\begin{pmatrix}
		C_0	\\
		C_1 \\
		\vdots \\
		C_N \\
		C
	\end{pmatrix} = \begin{pmatrix}
		\mathcal{O}(1) \\
		\mathcal{O}(\varepsilon)	\\
		\vdots \\
		\mathcal{O}(\varepsilon)	\\
		0
	\end{pmatrix}.
\end{equation}
Thus the appropriate scaling for the solution is $C_0=\mathcal{O}(1)$, $C_m=\mathcal{O}(\varepsilon)$ for $m=1,\ldots,N$, and $C=\mathcal{O}(\varepsilon)$. Using the scalings for $C_0,\ldots,C_N$, the components of the particular solution in Eq.~\eqref{eqn:kj_reg_particular_scaling}, and the eigenvectors in Eqs.~\eqref{eqn:kj_reg_evec_scaling1} and \eqref{eqn:kj_reg_evec_scaling2}, we find that the effective ingestion rates $k_j$ in Eq.~\eqref{eqn:effective_reaction_rate}  are $\mathcal{O}(1)$ for all $j=0,\ldots N-1$, which is the desired result of this appendix. 

\bibliographystyle{siamplain}
\bibliography{bibliography.bib}

\end{document}

%% file: shared.tex
\usepackage{lipsum}
\usepackage{amsfonts}
\usepackage{graphicx}
\usepackage{algorithmic}
\usepackage{amsmath,cite}
\usepackage{color}
\usepackage{caption}
\usepackage{subcaption}
\usepackage{cprotect}
\usepackage{epstopdf}
\usepackage{verbatim}
\usepackage[section]{placeins}
\usepackage{multirow}	
\usepackage{enumitem}  
\usepackage{IEEEtrantools}
\usepackage{multicol}
\usepackage{amsopn}
\usepackage{todonotes}
\ifpdf
  \DeclareGraphicsExtensions{.eps,.pdf,.png,.jpg}
\else
  \DeclareGraphicsExtensions{.eps}
\fi

\newsiamremark{remark}{Remark}
\newsiamremark{hypothesis}{Hypothesis}
\crefname{hypothesis}{Hypothesis}{Hypotheses}
\newsiamthm{claim}{Claim}
\newsiamremark{fact}{Fact}
\crefname{fact}{Fact}{Facts}

\headers{Coarse-Graining ABMs of Bacterial Infections}{Wesley J. M. Ridgway, and Raymond J. Spiteri}

\title{Coarse-Graining Agent-Based Models of Bacterial Infections}

\author{Wesley J.~M.~Ridgway\thanks{Department of Mathematics and Statistics, University of Saskatchewan, Saskatoon, Saskatchewan, Canada}
\and Raymond J. Spiteri\thanks{Department of Computer Science, University of Saskatchewan, Saskatoon, Saskatchewan, Canada.}}
